\documentclass[pra,superscriptaddress,floatfix,twocolumn,showpacs]{revtex4-2}
\usepackage{babel}
\usepackage{fmtcount}
\usepackage{verbatim}
\usepackage{amsmath}
\usepackage{amsfonts}
\usepackage{amssymb}
\usepackage{graphicx}
\usepackage{bm}
\usepackage{color}
\usepackage[dvipsnames]{xcolor}
\usepackage{hyperref}
\usepackage[active]{srcltx}
\usepackage{cases}
\usepackage{ulem}
\usepackage{multirow}
\usepackage{braket}
\usepackage[utf8]{inputenc}

\begin{document}

\title{Measurement-Induced Spectral Transition}

\author{Ken Mochizuki}
\affiliation{Department of Applied Physics, University of Tokyo, Tokyo 113-8656, Japan}
\affiliation{Nonequilibrium Quantum Statistical Mechanics RIKEN Hakubi Research Team, RIKEN Cluster for Pioneering Research (CPR), 2-1 Hirosawa, Wako 351-0198, Saitama, Japan}

\author{Ryusuke Hamazaki}
\affiliation{Nonequilibrium Quantum Statistical Mechanics RIKEN Hakubi Research Team, RIKEN Cluster for Pioneering Research (CPR), 2-1 Hirosawa, Wako 351-0198, Saitama, Japan}
\affiliation{RIKEN Interdisciplinary Theoretical and Mathematical Sciences Program (iTHEMS), 2-1 Hirosawa, Wako 351-0198, Saitama, Japan}

%\date{April 2023}
\begin{abstract}
We show that noisy quantum dynamics exposed to generalized measurements exhibit a spectral transition between gapless and gapped phases. 
To this end, we employ the Lyapunov spectrum obtained through singular values of a non-unitary matrix describing the dynamics. 
We discover that the gapless and gapped phases respectively correspond to the volume-law and area-law phases of the entanglement entropy for the dominant Lyapunov vector. 
This correspondence between the spectral gap and the scaling of entanglement offers an intriguing common structure with ground-state phase transitions. 
We also discuss some crucial differences from ground-state transitions, such as the extensive entanglement and the exponentially small gaps. 
Furthermore, we show that the spectral transition leads to the transition of the timescale for the memory loss of initial states. 
\end{abstract}

\maketitle

\paragraph*{Introduction.--}
In quantum systems described by time-independent generators, such as the Hamiltonians for isolated systems and the Liouvillians for dissipative systems, the generators' spectrum and gap provide essential information about the systems. 
For example, quantum phase transitions of ground states are accompanied by the gap closing of the Hamiltonian \cite{sachdev2001quantum,
hastings2005quasiadiabatic}, and the Liouvillian gap is related to the asymptotic decay rate towards the stationary state \cite{kessler2012dissipative,
cai2013algebraic,
bonnes2014dynamical,
znidaric2015relaxation,
minganti2018spectral,
mori2020resolving,
haga2021liouvillian}. 
For quantum phase transitions, ground states in gapless regimes exhibit behaviors qualitatively different from those in gapped phases in terms of the entanglement entropy and correlation functions \cite{holzhey1994geometric,
calabrese2004entanglement,
hastings2006spectral,
hastings2007area}.

The entanglement entropy, which captures intrinsic features of quantum states, 
has also attracted recent huge attention in noisy dynamics of monitored systems, where generators randomly depend on time. 
Owing to the competition between unitary dynamics and quantum measurements, where the former and latter respectively enhance and suppress the entanglement growth, measurement-induced entanglement transitions occur
\cite{li2018quantum,
cao2019entanglement,
chan2019unitary,
li2019measurement,
skinner2019measurement,
szyniszewski2019entanglement,
bao2020theory,
choi2020quantum,
fuji2020measurement,
gullans2020dynamical,
lunt2020measurement,
szyniszewski2020universality,
turkeshi2020measurement,
alberton2021entanglement,
lu2021spacetime,
agrawal2022entanglement,
barratt2022field,
block2022measurement,
minato2022fate,
muller2022measurement,
noel2022measurement,
sierant2022universal,
zabalo2022operator,
granet2023volume,
koh2023measurement,
le2023volume,
loio2023purification,
majidy2023critical,
oshima2023charge,
yamamoto2023localization,
kumar2024boundary,
aziz2024critical}. 
Many intriguing aspects of such measurement-induced transitions have been explored, including critical properties \cite{li2018quantum,
li2019measurement,
skinner2019measurement,
fuji2020measurement,
lunt2020measurement,
turkeshi2020measurement,
zabalo2022operator,
kumar2024boundary,
aziz2024critical} and timescales of relaxation \cite{gullans2020dynamical,
agrawal2022entanglement,
barratt2022field,
loio2023purification,
mochizuki2023distinguishability}.

Then, a natural question is whether the measurement-induced phase transitions can be understood as spectral transitions like the ground-state phase transitions in equilibrium. 
Unfortunately, this problem has been elusive due to the intrinsic randomness in the quantum measurement process.

\begin{figure}
\begin{center}
\includegraphics[width=9cm]{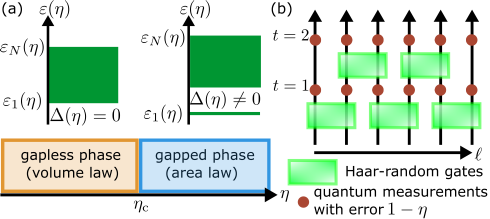}
\caption{(a) Schematic figure of the spectral transition. 
With varying a parameter $\eta$, the transition occurs between the gapless and gapped phases, where $\{\varepsilon_i(\eta)\}$ are the Lyapunov spectra of the time-evolution matrix. 
The threshold of the spectral transition coincides with that of the entanglement transition between the volume-law and area-law phases for the dominant Lyapunov vector corresponding to $\varepsilon_1(\eta)$. 
(b) Our model for the dynamics of qubits under measurements. 
After local unitary gates are applied, quantum measurements with errors $1-\eta$ are carried out for all sites. }
\label{fig:gap_dynamics}
\end{center}
\end{figure}

In this Letter, we discover a measurement-induced spectral transition leading to the entanglement transition like the ground-state phase transitions between gapless and gapped phases. 
While our monitored system lacks a time-independent generator, we focus on the Lyapunov spectrum obtained through singular values of a random non-unitary matrix describing the dynamics. 
As a result, we find a spectral transition from the gapless phase to the gapped phase, as shown in Fig. \ref{fig:gap_dynamics} (a). 
The gapless and gapped phases respectively correspond to the volume-law and area-law phases of the entanglement entropy for the dominant Lyapunov vector after long times. 
Such a correspondence between the entanglement scaling and the spectral gap is similar to the  entanglement transitions of ground states  
\cite{holzhey1994geometric,
calabrese2004entanglement,
hastings2006spectral,
hastings2007area}. 
Importantly, however, in the gapless phase, we find that scalings of the spectral gap and the entanglement entropy of monitored dynamics are distinct from those of ground states of well-known local Hamiltonians. 
We further argue that the spectral transition leads to the transition of the timescale at which observables become independent of initial states, reminiscent of the purification transition. 

%Thus, our study would provide one step toward understanding transitions in equilibrium and non-equilibrium quantum systems in a similar vein. 

\paragraph*{Setup.--}
We consider quantum dynamics depicted in Fig. \ref{fig:gap_dynamics} (b), where $L$ qubits are arrayed on a one-dimensional chain. 
The local unitary gates, which entangle nearest neighbor qubits, are given by $4\times4$ Haar-random unitary matrices. 
These gates are arranged in a brick-wall manner under the open boundary condition, where the unitary matrix at time $t$ is denoted as $U_t$. 
The generalized quantum measurements at time $t$, described by Kraus operators $M_t(\eta)=\bigotimes_{\ell=1}^L\mathsf{M}_{\omega_{t,\ell}}(\eta)$ with 
\begin{align}
    \mathsf{M}_\pm(\eta)=\frac{\sigma_0 \pm \eta\sigma_3}{\sqrt{2(1+\eta^2)}},
\end{align}
include the error $1-\eta$ where $0\leq\eta\leq 1$ \cite{supplemental-material_measurement-induced_spectral-transition}. 
Here, $\sigma_0$ is the $2\times2$ identity matrix, $\sigma_3$ is the $z$ component of the Pauli matrices, $\ell$ represents a position of a qubit, and $\omega_{t,\ell}$ takes $+$ or $-$, which correspond to the measurement outcomes for $\sigma_3^\ell$ under the error with $\sigma_3^\ell$ being the Pauli matrix at $\ell$. 
The probability distribution of $\{\omega_{t,\ell}\}$ is determined through the Born rule. 
The parameter $\eta$ describes the strength of the measurement, where $\eta=1$ and $\eta=0$ correspond to the projective measurement and no measurement, respectively. 
The dynamics from an initial pure state $\ket{\psi_0}$ become $\ket{\psi_t(\eta)}
=V_t(\eta)\ket{\psi_0}/\sqrt{\bra{\psi_0}V^\dagger_t(\eta)V_t(\eta)\ket{\psi_0}}$, 
where $V_t(\eta)=M_t(\eta)U_tM_{t-1}(\eta)U_{t-1} \cdots M_1(\eta)U_1$ implicitly depends on measurement outcomes $\{\omega_{t,\ell}\}$. 

We define an effective ``Hamiltonian" 
\begin{align}
    K_t(\eta)=-\frac{1}{2t}
    \ln\left[V_t(\eta)
    V_t^\dagger(\eta)\right],
\end{align} 
which leads to an analogy between ground-state phase transitions and measurement-induced transitions. 
The eigenvalues of $K_t(\eta)$, $\{\varepsilon_{t,i}(\eta,L)\}$, are arrayed as $\varepsilon_{t,i}(\eta,L)\leq\varepsilon_{t,i+1}(\eta,L)$ with $1 \leq i \leq N:=2^L$, where the singular values of $V_t(\eta)$ become $\Lambda_{t,i}(\eta,L)=e^{-\varepsilon_{t,i}(\eta,L)t}$. 
The $i$th Lyapunov vector $\ket{\Psi_{t,i}(\eta)}$ is the eigenmode of $K_t(\eta)$ whose eigenvalue is $\varepsilon_{t,i}(\eta,L)$. 
The dominant Lyapunov vector $\ket{\Psi_{t,1}(\eta)}$ is the ground state of the effective Hamiltonian. 
While we mainly explore pure-state dynamics, if we take the initial state as the maximally mixed state, the mixed state at $t$ becomes the Gibbs state with the Hamiltonian $K_t(\eta)$ and the inverse temperature $2t$. 
The partition function of the circuit~\cite{zabalo2022operator} is given by $\mathrm{tr}\left[e^{-2K_t(\eta)t}\right]$.

As an indicator for the measurement-induced phase transition, we focus on the spectral gap,
\begin{align}
    \Delta(\eta,L)
    =\varepsilon_2(\eta,L)-\varepsilon_1(\eta,L),
\end{align}
where $\varepsilon_i(\eta,L)=\lim_{t\rightarrow\infty}\varepsilon_{t,i}(\eta,L)$. 
When $t$ is sufficiently large such that $\Delta_t(\eta,L)=\varepsilon_{t,2}(\eta,L)-\varepsilon_{t,1}(\eta,L)\simeq\Delta(\eta,L)$, the spectral gap $\Delta(\eta,L)$ gives the relaxation time 
\begin{align}
\tau_\delta(\eta,L)
=\left|\frac{\ln(\delta)}
{\Delta(\eta,L)}\right|,
\end{align}
at which $\ket{\Psi_{t,i}(\eta)}$ with $i\geq2$ become negligible within the precision $\delta$ and thus $\ket{\psi_t(\eta)}\simeq\ket{\Psi_{t,1}(\eta)}$ is realized. 
The dominant Lyapunov vector 
is realized in the long-time regime irrespective of initial states, like stationary states in open quantum systems described by static generators. 
However, $\ket{\Psi_{t,1}(\eta)}$ randomly fluctuates in time, in contrast with stationary states which are independent of time. 

We note that, in several settings, the spectral gap has been computed at the transition point to extract critical properties of measurement-induced transitions \cite{zabalo2022operator,kumar2024boundary,aziz2024critical}. 
In contrast, we explore the scaling of $\Delta(\eta,L)$ and eigenstate properties in a wide parameter region of $\eta$.

\begin{figure*}
\begin{center}
\includegraphics[width=13.5cm]{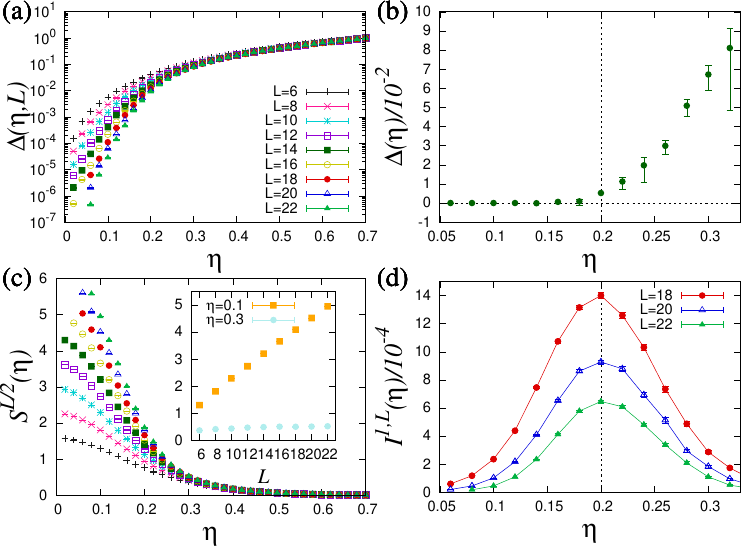}
\caption{(a) The spectral gap $\Delta(\eta,L)$ as functions of $\eta$ for various system sizes $L$. 
$\Delta(\eta,L)$ rapidly decreases with increasing $L$ for small $\eta$, while it is almost constant for large $\eta$. 
(b) The spectral gap $\Delta(\eta)$ in the thermodynamic limit. 
The spectral transition between the gapless and gapped phases occurs around $\eta_\mathrm{c}=0.2$. 
(c) The entanglement entropy $S^{L/2}(\eta)$ of the dominant Lyapunov vector $\ket{\Psi_{t,1}(\eta)}$ for various measurement strengths $\eta$ and system sizes $L$. 
The inset shows $S^{L/2}(\eta)$ as functions of the system size $L$, where the gapless and gapped phases exhibit
the volume-law (orange squares) and the area-law (blue circles) scalings of $S^{L/2}(\eta)$, respectively. 
(d) The mutual information of $\ket{\Psi_{t,1}(\eta)}$. 
A peak around $\eta_\mathrm{c}$ indicates the entanglement transition between the volume-law and area-law phases. 
The thresholds of the spectral transition and the entanglement transition coincide.}
\label{fig:gap_entanglement_e-dependence}
\end{center}
\end{figure*}

\begin{table*}[]
    \centering
    \begin{tabular}{|c|c|c|}\hline
          & ground-state transition in equilibrium
      &measurement-induced transition in noisy dynamics\\\hline
      gapped phase & 
      gap:\,$O(L^0)$,\ \ 
      entanglement entropy:\,$O(L^0)$ & gap:\,$O(L^0)$,\ \ 
      entanglement entropy:\,$O(L^0)$ \\\hline
      gapless phase & 
      gap:\,$O[1/\mathrm{poly}(L)]$,\ \ entanglement entropy:\,$O[\ln(L)]$
      & gap:\,$\exp[-O(L)]$,\ \ entanglement entropy:\,$O(L)$
      \\\hline
    \end{tabular}
    \caption{Comparison between ground-state transitions of typical one-dimensional isolated quantum systems described by local Hamiltonians and the measurement-induced transition explored in this work, in terms of the spectral gap and the half-chain entanglement entropy. 
    Behaviors in gapped phases are qualitatively similar, while those in gapless phases are distinct. }
    \label{tab:comparison_equilibrium-non-equilibrium}
\end{table*}

\paragraph*{Transition of the spectral gap.--}
Our monitored dynamics exhibit a transition in terms of the Lyapunov spectrum. 
To compute the Lyapunov spectral gap, we employ the method developed to analyze classical chaotic systems \cite{ershov1998concept}.  
In essence, we prepare randomly chosen vectors and evaluate decay rates of the volume formed by those vectors using the Gram-Schmidt orthonormalization. 
We use a sampling method based on one trajectory and take the time average of the spectral gap in the long-time regime. 
The details of the numerical calculations are given in Supplemental Material \cite{supplemental-material_measurement-induced_spectral-transition}. 

Figure \ref{fig:gap_entanglement_e-dependence} (a) shows $\Delta(\eta,L)$ for various $\eta$ and $L$. 
We can understand that $\Delta(\eta,L)$ rapidly decreases with increasing $L$ for small $\eta$, which indicates $\Delta(\eta)=0$, where
\begin{align}
    \Delta(\eta)
    =\lim_{L\rightarrow\infty}
    \Delta(\eta,L).
\end{align}
In contrast, $\Delta(\eta,L)$ is almost independent of $L$ for large $\eta$, which indicates $\Delta(\eta)\neq0$. 
To compute $\Delta(\eta)$, we extrapolate $\Delta(\eta,L)$ using the fitting function $\Delta_\mathrm{fit}(\eta,L)=\Delta(\eta)+\alpha(\eta)[\beta(\eta)]^{-L}$. 
Through the least-square fitting with $10 \leq L \leq 22$, we obtain Fig. \ref{fig:gap_entanglement_e-dependence} (b), which clearly shows the transition between the gapless phase with $\Delta(\eta)=0$ and the gapped phase with $\Delta(\eta)\neq0$. 
Indeed, for $\eta<\eta_\mathrm{c}=0.2$ ($\eta\geq\eta_\mathrm{c}$), we see that the bottom values of the error bars are smaller (larger) than $0$, indicating that the system is in the gapless (gapped) phase. 
The spectral transition originates from the competition between unitary dynamics and quantum measurements since the transition vanishes when there is no unitary gate \cite{supplemental-material_measurement-induced_spectral-transition}. 
We note that Refs. \cite{de2023universality,bulchandani2024random} explore Lyapunov exponents and purification times of dynamics by non-local matrices, where no transition is found. 
This indicates that the locality of $U_t$ and $M_t(\eta)$ plays a key role in measurement-induced transitions~\cite{block2022measurement,
minato2022fate}.

Ground states in isolated quantum systems also exhibit the gapless-gapped transition. 
Table \ref{tab:comparison_equilibrium-non-equilibrium} compares the behaviors of the gaps in our monitored dynamics and typical isolated systems described by local Hamiltonians. 
While these are qualitatively similar in gapped phases, there is a qualitative difference in gapless phases; our monitored dynamics exhibit exponentially small gaps in contrast to polynomially small gaps of local Hamiltonians in isolated quantum systems \cite{zeng2019quantum}.

\paragraph*{Transition of the entanglement.--}
We show that the above spectral transition coincides with the entanglement transition. 
We present the entanglement transition for the dominant Lyapunov vector $\ket{\Psi_{t,1}(\eta)}$ to make a clear comparison with ground-state phase transitions. 
Note that the correspondence also applies to the entanglement transition for states after $t\propto L$, often discussed in previous works \cite{skinner2019measurement}, while the data is not shown.

We show that the critical point of the spectral transition coincides with that of the entanglement entropy for $\ket{\Psi_{t,1}(\eta)}$. 
To this end, we compute the entanglement entropy $S^A(\eta)=\frac{1}{T}\sum_{t=\tau+1}^{\tau+T}S_t^A(\eta)$ where $S_t^A(\eta)=-\mathrm{tr}_A\left(\rho_t^A(\eta)\ln[\rho_t^A(\eta)]\right)$ with $\rho_t^A(\eta)=\mathrm{tr}_{\overline{A}}\left(\ket{\psi_t(\eta)}\bra{\psi_t(\eta)}\right)$, averaged over $T=10^4$ time steps after $\tau\geq\tau_\delta(\eta,L)$ with $\delta=10^{-2}$. 
Here, $\mathrm{tr}_{\overline{A}}$ is the partial trace with respect to the complement of the subsystem $A$. 
Figure \ref{fig:gap_entanglement_e-dependence} (c) shows $S^{L/2}(\eta)$ for various $\eta$ and $L$. 
We can see that the entanglement entropy of the half chain exhibits the volume-law scaling, $S^{L/2}(\eta) \propto L$, for small $\eta$. 
On the other hand, the entanglement entropy exhibits the area-law scaling $S^{L/2}(\eta) \propto L^0$ for large $\eta$.

It is known that the transition point between the volume-law and area-law phases exhibits the peak of the mutual information $I_t^{A,B}(\eta)=S_t^A(\eta)+S_t^B(\eta)-S_t^{AB}(\eta)$
%since it gives the upper bound of correlation functions 
\cite{walf2008area,li2019measurement}. 
Figure \ref{fig:gap_entanglement_e-dependence} (d) shows the mutual information of $\ket{\psi_t(\eta)}\simeq\ket{\Psi_{t,1}(\eta)}$ averaged over $T=10^5$ time steps, $I^{A,B}(\eta)=\frac{1}{T}\sum_{t=\tau+1}^{\tau+T}I_t^{A,B}(\eta)$, where $A$ and $B$ are qubits at $\ell=1$ and $\ell=L$, respectively. 
We can see that $I^{1,L}(\eta)$ exhibits a peak around $\eta_\mathrm{c}=0.2$, which coincides with the spectral transition. 
Note that a peak of the mutual information was also reported in Ref. \cite{li2019measurement} with a different boundary condition. 

The coincidence of the spectral transition and the entanglement transition is analogous to ground-state phase transitions between gapless and gapped phases in equilibrium \cite{holzhey1994geometric,
calabrese2004entanglement,
hastings2006spectral,
hastings2007area}. 
This suggests that the relation between the spectral gap and the scaling of the entanglement also applies to measurement-induced phenomena, as well as the equilibrium physics. 
In particular, behaviors of the entanglement entropy in gapped phases are qualitatively similar, as listed in Table \ref{tab:comparison_equilibrium-non-equilibrium}. 
Meanwhile, a crucial distinction arises in the gapless phases; the entanglement exhibits the volume-law scaling for $\ket{\Psi_{t,1}(\eta)}$ of the monitored dynamics, while the logarithmic scaling appears for
the ground states of typical isolated systems described by local Hamiltonians \cite{eisert2010colloquium}. 

To clarify the distinction between the gapless phases in the two transitions, we explore the locality of $K_t(\eta)$. 
As shown in End Matter, we find that $K_t(\eta)$ includes long-range and many-body interactions for small $\eta$, suggesting that such non-locality is the origin of the exponentially small gap and the volume-law entanglement in the gapless phase for the monitored dynamics. 
We note that gapless phases in equilibrium exhibit criticality, while critical behaviors of monitored dynamics are observed at the transition point \cite{li2018quantum,
li2019measurement,
skinner2019measurement,
fuji2020measurement,
lunt2020measurement,
turkeshi2020measurement,
zabalo2022operator,
kumar2024boundary,
aziz2024critical}, which may also be the key to understanding the distinction in gapless phases.

\begin{figure}
\begin{center}
\includegraphics[width=7cm]{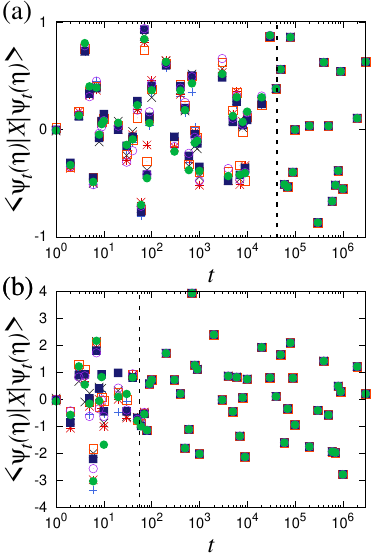}
\caption{Expectation values of $X=\sum_\ell\sigma_1^\ell$ with (a) $\eta=0.1$ in the gapless phase and (b) $\eta=0.3$ in the gapped phase. 
The black broken lines represent $\tau_\delta(\eta,L)=|\ln(\delta)/\Delta(\eta,L)|$ with $\delta=10^{-2}$. 
Here, different symbols correspond to randomly chosen different initial states, and the same $V_t(\eta)$ is applied to all initial states. 
The expectation values become almost the same after $\tau_\delta(\eta,L)$, and the timescale in the gapless phase is much longer than that in the gapped phase. 
The system size is $L=18$ in both (a) and (b).}
\label{fig:observable_t-dependence}
\end{center}
\end{figure}

\paragraph*{Memory loss effect.--}
The spectral transition also leads to the transition of the timescale for the memory loss, i.e., $\tau_\delta(\eta,L) \simeq \exp[O(L)]$ in the gapless phase and  $\tau_\delta(\eta,L)\simeq O(L^0)$ in the gapped phase. 
Indeed, if we fix $V_t(\eta)$, $\ket{\psi_t(\eta)}\simeq\ket{\Psi_{t,1}(\eta)}$ is independent of initial states $\ket{\psi_0}$ for $t\geq\tau_\delta(\eta,L)$ within the precision $\delta$. 
Figure \ref{fig:observable_t-dependence} shows $\bra{\psi_t(\eta)}X\ket{\psi_t(\eta)}$ with $X=\sum_\ell\sigma_1^\ell$ for several trajectories starting from different initial states, where $\sigma_1^\ell=\left(\bigotimes_{m=1}^{\ell-1}\sigma_0\right)\otimes\sigma_1\otimes\left(\bigotimes_{m=\ell+1}^L\sigma_0\right)$ is the $x$ component of the Pauli matrices at the site $\ell$. 
Here, $V_t(\eta)$ is generated through the Born rule based on one trajectory, and the same $V_t(\eta)$ is also applied to other initial states. 
We can see that the expectation values corresponding to different initial states take almost the same value when $t\geq\tau_\delta(\eta,L)$ with $\delta=10^{-2}$, while different initial states lead to different expectation values when $t\ll\tau_\delta(\eta,L)$. 
We can understand that the timescale $\tau_\delta(\eta,L)$ of the memory loss effect for $\eta<\eta_\mathrm{c}$ is much larger than that for $\eta>\eta_\mathrm{c}$, due to the spectral transition. 

We argue that the spectral gap, which becomes independent of trajectories in our monitored dynamics of pure states \cite{supplemental-material_measurement-induced_spectral-transition}, takes the same value even when the initial state is a mixed state. 
Indeed, we can confirm that the dynamics starting from the maximally mixed state exhibit almost the same spectral gap as that from randomly chosen pure states, as shown in End Matter. 
This argument indicates that the spectral transition and thus the entanglement transition in the monitored dynamics correspond to the purification transition: the volume-law (area-law) phase exhibits an exponentially small (constant) rate with respect to the system size for the initial mixed state to become a pure state approximately. 
This is consistent with the purification transition exposed to projective measurements \cite{gullans2020dynamical}.

\paragraph*{Structures of the entire spectrum.--}
While we have focused on the spectral gap, we next compare the entire spectra of our effective Hamiltonian and Hamiltonians in one-dimensional systems. 
Spectra of locally interacting many-body Hamiltonians usually have the following two structures: (a) the width of the spectrum scales at most linearly with $L$, which means that the typical level spacing is $\exp[-O(L)]$, and (b) the spectral gap becomes $O[1/\mathrm{poly}(L)]$ in gapless regimes or constant above a few states exponentially close to the ground state in gapped phases. 
We can analytically show that $\{\varepsilon_i(\eta)\}$ of our monitored dynamics also satisfy (a).
Indeed, employing the majorization, the Horn theorem, and properties of doubly stochastic matrices \cite{hiai2024log}, we can obtain \cite{supplemental-material_measurement-induced_spectral-transition}
\begin{align}
    \varepsilon_N(\eta,L)-\varepsilon_1(\eta,L)\leq 
    L\ln\left(\frac{1+\eta}{1-\eta}\right).
    \label{eq:inequality_width}
\end{align}
On the other hand, the monitored dynamics do not satisfy (b) in the gapless phase; the Lyapunov spectra of low-lying excited states become $\exp[-O(L)]$, which is numerically illustrated in Supplemental Material \cite{supplemental-material_measurement-induced_spectral-transition}.

\paragraph*{Conclusion.--}
We have shown that the quantum dynamics exposed to generalized measurements exhibit the spectral transition, whose threshold coincides with that of the entanglement transition. 
We have also shown that the spectral transition leads to the transition of the memory-loss timescale of pure states and argued that it also leads to the purification transition of mixed states. 

Our results suggest that Lyapunov analysis can reveal different measurement-induced phases in analogy with equilibrium quantum phase transitions. 
It is thus an intriguing future work to explore other counterparts of equilibrium phase transitions, such as topological transitions and spontaneous symmetry breaking, in monitored dynamics via Lyapunov analysis. 
It is also important to study the relation between our effective Hamiltonian and the Hamiltonians in replica methods \cite{bao2020theory,
muller2022measurement,
majidy2023critical}.
%This is because the coincidence of the spectral and entanglement transitions, common with ground-state phase transitions between gapless and gapped phases in equilibrium, suggests that monitored dynamics can exhibit phenomena analogous to equilibrium phase transitions. 
%Meanwhile, in gapless phases, the scalings of the entanglement and the gap in the monitored dynamics are qualitatively different from those in well-known isolated quantum systems described by local Hamiltonians. 
%Entanglement scaling of ground states is strongly affected by the locality of Hamiltonians.  
%On the other hand, it is non-trivial whether $K_t(\eta)$ becomes local. 
%Thus, $K_t(\eta)$ can be a non-local ``Hamiltonian," which may be the origin for the scalings of the spectral gap and entanglement in measurement-induced transitions distinct from those in ground-state phase transitions. 

\begin{acknowledgements}
We thank Yohei Fuji, Hisanori Oshima, Xhek Turkeshi, Marco Schir\'o, and Keiji Saito for fruitful discussions, Nobuyuki Yoshioka and Toshihiro Yada for valuable comments on the manuscript, and Zongping Gong for both. 
Derivation of the spectral gap under no unitary gate given in Supplemental Material is partly based on a note written by Toshihiro Yada. 
This work was supported by JST ERATO Grant Number JPMJER2302, Japan. 
K.M. was supported by JSPS KAKENHI Grant. No. JP23K13037. 
R.H. was supported by JSPS KAKENHI Grant No. JP24K16982.
\end{acknowledgements}

\onecolumngrid
\newpage
\section*{End Matter}
\twocolumngrid
%\appendix

\section{Non-local interactions of the effective Hamiltonian in the gapless phase}
\label{sec:interaction}
We explore whether $K_t(\eta)$ includes non-local terms to compare effective Hamiltonians in our monitored dynamics and Hamiltonians describing equilibrium quantum systems. 
This is because Hamiltonians in the latter case typically have only local and few-body interactions, such as nearest-neighbor interactions, and have no or weak long-range interactions while such locality is not guaranteed for our effective Hamiltonian $K_t(\eta)$. 
If $K_t(\eta)$ in the monitored dynamics includes long-range interactions that are not negligible, the non-locality of $K_t(\eta)$ can be the origin for the exponentially small gaps and extensive entanglement of ground states, since such behaviors rarely emerge in systems described by local Hamiltonians. 

As detailed in Supplemental Material \cite{supplemental-material_measurement-induced_spectral-transition}, we compute the Lyapunov spectrum by evaluating the decay rates of Gram-Schmidt vectors repeatedly at every $t=sb$ time step with $s$ and $b$ being positive integers.  
Here, $b$ is a fixed time interval, and $s$ increases to $s+1$ as $\ket{\psi_{sb}(\eta)}$ experiences $b$-step time evolution and becomes $\ket{\psi_{(s+1)b}(\eta)}$. 
If the measurement strength $\eta$ is small and the system size $L$ is not large, we can obtain the effective Hamiltonian $K_t(\eta)$ from the entire Lyapunov spectrum $\{\varepsilon_{t,i}(\eta,L)\}$ and Gram-Schmidt vectors based on small $b$. 
This is because the value of $\exp\left[-\varepsilon_{t,i}(\eta,L)b\right]$ for large $i \simeq N$ can be within the numerical precision when the spectral width $\varepsilon_{t,N}(\eta,L)-\varepsilon_{t,1}(\eta,L)$ is small enough with small $\eta$ and $L$. 
If $\eta$ and/or $L$ are large, $\exp\left[-\varepsilon_{t,i}(\eta,L)b\right]$ can easily deviate from the numerical precision. 
The details of numerics are given in Supplemental Material \cite{supplemental-material_measurement-induced_spectral-transition}.

We focus on the Pauli strings of distance $r$,
\begin{align}
    P_i^r=\prod_{\ell=1}^{r+1}
    \sigma_i^\ell,
\end{align}
where $i=1,2,3$. 
We compute the absolute values of coefficients for $P_i^r$ included in the effective Hamiltonian, 
\begin{align}
    W_{t,i}^r(\eta)=\frac{1}{c}\sum_{\tilde{c}=0}^{c-1}
    \frac{\left|\mathrm{tr}[P_i^rK_{t-\tilde{c}b}(\eta)]\right|}{N},
\end{align}
averaged over $c$ steps. 
Note that non-negligible weights of $P_i^r$ for $r \sim L$ indicate that the effective Hamiltonian includes non-local and many-body components. 

\begin{figure}
\begin{center}
\includegraphics[width=7cm]{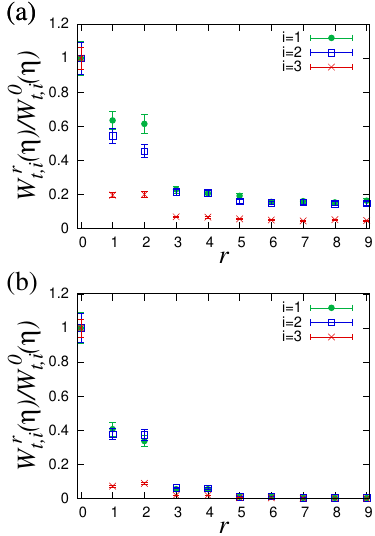}
\caption{The time average of the absolute values of the coefficients of $P_i^r$ that are included in $K_t(\eta)$ with $L=10$ and $c=100$. 
(a) The measurement strength is $\eta=0.01$ and the time step is $t=sb$ with $s=6853$ and $b=2048$. 
(b) The measurement strength is $\eta=0.4$ and the time step is $t=sb$ with $s=38546$ and $b=2$. 
Note that we use smaller $t=sb$ for larger $\eta$ because the convergence of the Lyapunov spectrum becomes faster as $\eta$ is increased.}
\label{fig:amplitude_r-dependence}
\end{center}
\end{figure}

Figure \ref{fig:amplitude_r-dependence} (a) with small $\eta$ in the gapless phase suggests that $W_{t,i}^r(\eta)$ converges to a constant value even for large $r$; the effective Hamiltonians $K_t(\eta)$ include long-range many-body interactions whose magnitudes are not negligible. 
In contrast, as shown Fig. \ref{fig:amplitude_r-dependence} (b), with larger $\eta=0.4$ at which the gap is open, $W_{t,i}^r(\eta)$ exhibits much faster decay as $r$ is increased. 
This suggests that long-range many-body interactions in $K_t(\eta)$ are negligible. 
We note that, due to finite-size effects, it is difficult to determine the functional form for the decay of $W_{t,i}(\eta)$, e.g. exponential or polynomial decay, through the direct construction of effective Hamiltonians. 
This is because it takes huge numerical costs to compute the entire Lyapunov spectra and Lyapunov vectors. 
Exploring $W_{t,i}(\eta)$ in larger systems by developing efficient numerical methods, such as using matrix product state representations, should be an important future work. 
%This observation is consistent with the fact that ground states of local Hamiltonians exhibit the area-law scaling of the entanglement entropy in the gapped phase \cite{eisert2010colloquium,zeng2019quantum}. 

In the gapless phases of the equilibrium quantum systems described by local Hamiltonians, it is usually observed that the spectral gaps become polynomially small and that ground states exhibit the logarithmic scaling of the entanglement entropy with respect to the system size \cite{eisert2010colloquium,zeng2019quantum}. 
Exponentially small gaps and ground states with volume-law entanglement \cite{zeng2019quantum} are rarely found in equilibrium quantum systems described by local Hamiltonians, though they are not strictly prohibited~\cite{gottesman2010entanglement,
irani2010ground,
vitagliano2010volume,
ramirez2014conformal,
udagawa2017finite,
zhang2017novel}. 
On the other hand, in the gapless phase of our monitored dynamics, the effective Hamiltonians $K_t(\eta)$ exhibit exponentially small spectral gaps and the volume-law entanglement of ground states. 
Our results suggest that the non-local many-body interactions in the effective Hamiltonian $K_t(\eta)$ of monitored dynamics should be the origin of the behaviors distinct from those in equilibrium systems. 
%Thus, non-local interactions in $K_t(\eta)$, suggested by Fig. \ref{fig:amplitude_r-dependence} (a), should be the origin of the difference in the scalings of the spectral gap and entanglement in gapless phases of these two transitions. 

\begin{figure}[tbp]
\begin{center}
\includegraphics[width=7.5cm]{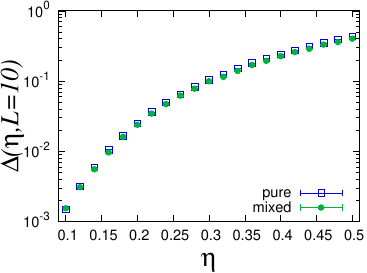}
\caption{The spectral gap $\Delta(\eta,L)$ as functions of $\eta$. 
Blue squares are obtained through the sampling method based on one pure-state trajectory, in the long-time regime after $\tau\geq\tau_\delta(\eta,L)=\left|\ln(\delta)/\Delta(\eta,L)\right|$ with $\delta=10^{-2}$. 
Green circles are obtained through the diagonalization method based on $500$ trajectories of mixed-state dynamics. 
While the latter result is evaluated in a short-time regime up to $t=500$ due to computational limitations, it is already almost equal to the result for the long-time pure-state dynamics.}
\label{fig:gap_pure-mixed_e-dependence}
\end{center}
\end{figure}

\section{The spectral gap when the initial state is the maximally mixed state}
\label{sec:mixed}
The spectral gap $\Delta(\eta,L)$ for an initial maximally mixed state should give the timescale of purification; the state approximately becomes a pure state with timescale proportional to $1/\Delta(\eta,L)$. 
To show this, we consider the dynamics of mixed states described by density matrices $\rho_t(\eta)$, while the dynamics of pure states are explored in the main text. 
When local Haar-random unitaries are applied, the density matrix is transformed as
\begin{align}
    \rho_t(\eta)\rightarrow
    U_t\rho_t(\eta)U_t^\dagger.
\end{align}
After that, generalized measurements for all sites are carried out. 
When the measurement is done at a position $\ell$, the probability that the outcome becomes $\pm$ is
\begin{align}
    p_{t,\ell}(\pm,\eta)=\mathrm{tr}[M_{\pm,\ell}(\eta)\rho_t(\eta)M_{\pm,\ell}^\dagger(\eta)],
\end{align}
where $M_{\pm,\ell}(\eta)=\left(\bigotimes_{m=1}^{\ell-1}\sigma_0\right)\otimes\mathsf{M}_{\pm}(\eta)\otimes\left(\bigotimes_{m=\ell+1}^L\sigma_0\right)$. 
After the measurement at $\ell$, probabilistically depending on the measurement outcome $\pm$, the density matrix becomes
\begin{align}
    \rho_t(\eta)\rightarrow
    \frac{M_{\pm,\ell}(\eta)\rho_t(\eta)M_{\pm,\ell}^\dagger(\eta)}
    {p_{t,\ell}(\pm,\eta)}.
\end{align}
To evaluate the purification timescale, we choose the initial state as the maximally mixed state,
\begin{align}
    \rho_0=I_N/N,
\end{align}
where $I_N$ is the $N \times N$ identity matrix. 
Then, through the dynamics by Haar-random unitary circuits and generalized measurements, the density matrix becomes
\begin{align}
    \rho_t(\eta)=\frac{V_t(\eta)V_t^\dagger(\eta)}
    {\mathrm{tr}[V_t(\eta)V_t^\dagger(\eta)]}.
\end{align}
If $\Lambda_{t,2}^2(\eta)/\Lambda_{t,1}^2(\eta)\simeq\exp[-2\Delta(\eta,L)t]$ becomes quite smaller than $1$, the matrix rank of $\rho_t(\eta)$ can be regarded as $1$ approximately, and thus $\rho_t(\eta)$ can be approximated as the pure ground state,
\begin{align}
    \rho_t(\eta)\simeq\ket{\Psi_{t,1}(\eta)}\bra{\Psi_{t,1}(\eta)}.
\end{align}
Therefore, $1/\Delta(\eta,L)$ becomes proportional to the timescale for the purification.

Now, let us compare $\Delta(\eta,L)$ for mixed states with that for pure states. 
Figure \ref{fig:gap_pure-mixed_e-dependence} shows the spectral gap computed through dynamics starting from a randomly chosen initial pure state (blue squares) and that from the maximally mixed initial state (green circles) in a small system with $L=10$. 
In the latter case, the Lyapunov gap is computed through diagonalizing $\rho_t(\eta)$ in the short-time regime $20 \leq t \leq 500$ with the ensemble average over $500$ trajectories because of computational limitations. 
From Fig. \ref{fig:gap_pure-mixed_e-dependence}, we can understand that the dynamics of pure and mixed states give almost the same spectral gap. 
We assume that this coincidence of the spectral gap persists in larger systems. 
Thus, the spectral transition found in the main text indicates the transition of the purification timescale.

While the purification transition has been reported in quantum dynamics exposed to projective measurements \cite{gullans2020dynamical}, we have explored purification dynamics exposed to generalized measurements.  
Exploring the discrepancy and similarity between projective and generalized measurements is an interesting future work. 
This is because $\{\Lambda_{t,i}(\eta\neq1,L)\}$ do not become zero through generalized measurements, while singular values of time-evolution operators can be strictly zero and the Lyapunov spectrum with $i>1$ can be ill-defined through projective measurements. 
This fact may make relaxation dynamics under generalized measurements distinct from those under projective measurements.

%\newpage
\bibliographystyle{apsrev4-2}
\bibliography{reference.bib}

\begin{thebibliography}{62}%
\makeatletter
\providecommand \@ifxundefined [1]{%
 \@ifx{#1\undefined}
}%
\providecommand \@ifnum [1]{%
 \ifnum #1\expandafter \@firstoftwo
 \else \expandafter \@secondoftwo
 \fi
}%
\providecommand \@ifx [1]{%
 \ifx #1\expandafter \@firstoftwo
 \else \expandafter \@secondoftwo
 \fi
}%
\providecommand \natexlab [1]{#1}%
\providecommand \enquote  [1]{``#1''}%
\providecommand \bibnamefont  [1]{#1}%
\providecommand \bibfnamefont [1]{#1}%
\providecommand \citenamefont [1]{#1}%
\providecommand \href@noop [0]{\@secondoftwo}%
\providecommand \href [0]{\begingroup \@sanitize@url \@href}%
\providecommand \@href[1]{\@@startlink{#1}\@@href}%
\providecommand \@@href[1]{\endgroup#1\@@endlink}%
\providecommand \@sanitize@url [0]{\catcode `\\12\catcode `\$12\catcode
  `\&12\catcode `\#12\catcode `\^12\catcode `\_12\catcode `\%12\relax}%
\providecommand \@@startlink[1]{}%
\providecommand \@@endlink[0]{}%
\providecommand \url  [0]{\begingroup\@sanitize@url \@url }%
\providecommand \@url [1]{\endgroup\@href {#1}{\urlprefix }}%
\providecommand \urlprefix  [0]{URL }%
\providecommand \Eprint [0]{\href }%
\providecommand \doibase [0]{https://doi.org/}%
\providecommand \selectlanguage [0]{\@gobble}%
\providecommand \bibinfo  [0]{\@secondoftwo}%
\providecommand \bibfield  [0]{\@secondoftwo}%
\providecommand \translation [1]{[#1]}%
\providecommand \BibitemOpen [0]{}%
\providecommand \bibitemStop [0]{}%
\providecommand \bibitemNoStop [0]{.\EOS\space}%
\providecommand \EOS [0]{\spacefactor3000\relax}%
\providecommand \BibitemShut  [1]{\csname bibitem#1\endcsname}%
\let\auto@bib@innerbib\@empty
%</preamble>
\bibitem [{\citenamefont {Sachdev}(2001)}]{sachdev2001quantum}%
  \BibitemOpen
  \bibfield  {author} {\bibinfo {author} {\bibfnamefont {S.}~\bibnamefont
  {Sachdev}},\ }\href@noop {} {\emph {\bibinfo {title} {Quantum Phase
  Transitions}}}\ (\bibinfo  {publisher} {Cambridge University Press,
  Cambridge, England},\ \bibinfo {year} {2001})\BibitemShut {NoStop}%
\bibitem [{\citenamefont {Hastings}\ and\ \citenamefont
  {Wen}(2005)}]{hastings2005quasiadiabatic}%
  \BibitemOpen
  \bibfield  {author} {\bibinfo {author} {\bibfnamefont {M.~B.}\ \bibnamefont
  {Hastings}}\ and\ \bibinfo {author} {\bibfnamefont {X.-G.}\ \bibnamefont
  {Wen}},\ }\bibfield  {title} {\bibinfo {title} {Quasiadiabatic continuation
  of quantum states: The stability of topological ground-state degeneracy and
  emergent gauge invariance},\ }\href
  {https://doi.org/10.1103/PhysRevB.72.045141} {\bibfield  {journal} {\bibinfo
  {journal} {Phys. Rev. B}\ }\textbf {\bibinfo {volume} {72}},\ \bibinfo
  {pages} {045141} (\bibinfo {year} {2005})}\BibitemShut {NoStop}%
\bibitem [{\citenamefont {Kessler}\ \emph {et~al.}(2012)\citenamefont
  {Kessler}, \citenamefont {Giedke}, \citenamefont {Imamoglu}, \citenamefont
  {Yelin}, \citenamefont {Lukin},\ and\ \citenamefont
  {Cirac}}]{kessler2012dissipative}%
  \BibitemOpen
  \bibfield  {author} {\bibinfo {author} {\bibfnamefont {E.~M.}\ \bibnamefont
  {Kessler}}, \bibinfo {author} {\bibfnamefont {G.}~\bibnamefont {Giedke}},
  \bibinfo {author} {\bibfnamefont {A.}~\bibnamefont {Imamoglu}}, \bibinfo
  {author} {\bibfnamefont {S.~F.}\ \bibnamefont {Yelin}}, \bibinfo {author}
  {\bibfnamefont {M.~D.}\ \bibnamefont {Lukin}},\ and\ \bibinfo {author}
  {\bibfnamefont {J.~I.}\ \bibnamefont {Cirac}},\ }\bibfield  {title} {\bibinfo
  {title} {Dissipative phase transition in a central spin system},\ }\href
  {https://doi.org/10.1103/PhysRevA.86.012116} {\bibfield  {journal} {\bibinfo
  {journal} {Phys. Rev. A}\ }\textbf {\bibinfo {volume} {86}},\ \bibinfo
  {pages} {012116} (\bibinfo {year} {2012})}\BibitemShut {NoStop}%
\bibitem [{\citenamefont {Cai}\ and\ \citenamefont
  {Barthel}(2013)}]{cai2013algebraic}%
  \BibitemOpen
  \bibfield  {author} {\bibinfo {author} {\bibfnamefont {Z.}~\bibnamefont
  {Cai}}\ and\ \bibinfo {author} {\bibfnamefont {T.}~\bibnamefont {Barthel}},\
  }\bibfield  {title} {\bibinfo {title} {Algebraic versus exponential
  decoherence in dissipative many-particle systems},\ }\href
  {https://doi.org/10.1103/PhysRevLett.111.150403} {\bibfield  {journal}
  {\bibinfo  {journal} {Phys. Rev. Lett.}\ }\textbf {\bibinfo {volume} {111}},\
  \bibinfo {pages} {150403} (\bibinfo {year} {2013})}\BibitemShut {NoStop}%
\bibitem [{\citenamefont {Bonnes}\ \emph {et~al.}(2014)\citenamefont {Bonnes},
  \citenamefont {Charrier},\ and\ \citenamefont
  {L\"auchli}}]{bonnes2014dynamical}%
  \BibitemOpen
  \bibfield  {author} {\bibinfo {author} {\bibfnamefont {L.}~\bibnamefont
  {Bonnes}}, \bibinfo {author} {\bibfnamefont {D.}~\bibnamefont {Charrier}},\
  and\ \bibinfo {author} {\bibfnamefont {A.~M.}\ \bibnamefont {L\"auchli}},\
  }\bibfield  {title} {\bibinfo {title} {Dynamical and steady-state properties
  of a bose-hubbard chain with bond dissipation: A study based on matrix
  product operators},\ }\href {https://doi.org/10.1103/PhysRevA.90.033612}
  {\bibfield  {journal} {\bibinfo  {journal} {Phys. Rev. A}\ }\textbf {\bibinfo
  {volume} {90}},\ \bibinfo {pages} {033612} (\bibinfo {year}
  {2014})}\BibitemShut {NoStop}%
\bibitem [{\citenamefont {\ifmmode \check{Z}\else
  \v{Z}\fi{}nidari\ifmmode~\check{c}\else
  \v{c}\fi{}}(2015)}]{znidaric2015relaxation}%
  \BibitemOpen
  \bibfield  {author} {\bibinfo {author} {\bibfnamefont {M.}~\bibnamefont
  {\ifmmode \check{Z}\else \v{Z}\fi{}nidari\ifmmode~\check{c}\else
  \v{c}\fi{}}},\ }\bibfield  {title} {\bibinfo {title} {Relaxation times of
  dissipative many-body quantum systems},\ }\href
  {https://doi.org/10.1103/PhysRevE.92.042143} {\bibfield  {journal} {\bibinfo
  {journal} {Phys. Rev. E}\ }\textbf {\bibinfo {volume} {92}},\ \bibinfo
  {pages} {042143} (\bibinfo {year} {2015})}\BibitemShut {NoStop}%
\bibitem [{\citenamefont {Minganti}\ \emph {et~al.}(2018)\citenamefont
  {Minganti}, \citenamefont {Biella}, \citenamefont {Bartolo},\ and\
  \citenamefont {Ciuti}}]{minganti2018spectral}%
  \BibitemOpen
  \bibfield  {author} {\bibinfo {author} {\bibfnamefont {F.}~\bibnamefont
  {Minganti}}, \bibinfo {author} {\bibfnamefont {A.}~\bibnamefont {Biella}},
  \bibinfo {author} {\bibfnamefont {N.}~\bibnamefont {Bartolo}},\ and\ \bibinfo
  {author} {\bibfnamefont {C.}~\bibnamefont {Ciuti}},\ }\bibfield  {title}
  {\bibinfo {title} {Spectral theory of liouvillians for dissipative phase
  transitions},\ }\href {https://doi.org/10.1103/PhysRevA.98.042118} {\bibfield
   {journal} {\bibinfo  {journal} {Phys. Rev. A}\ }\textbf {\bibinfo {volume}
  {98}},\ \bibinfo {pages} {042118} (\bibinfo {year} {2018})}\BibitemShut
  {NoStop}%
\bibitem [{\citenamefont {Mori}\ and\ \citenamefont
  {Shirai}(2020)}]{mori2020resolving}%
  \BibitemOpen
  \bibfield  {author} {\bibinfo {author} {\bibfnamefont {T.}~\bibnamefont
  {Mori}}\ and\ \bibinfo {author} {\bibfnamefont {T.}~\bibnamefont {Shirai}},\
  }\bibfield  {title} {\bibinfo {title} {Resolving a discrepancy between
  liouvillian gap and relaxation time in boundary-dissipated quantum many-body
  systems},\ }\href {https://doi.org/10.1103/PhysRevLett.125.230604} {\bibfield
   {journal} {\bibinfo  {journal} {Phys. Rev. Lett.}\ }\textbf {\bibinfo
  {volume} {125}},\ \bibinfo {pages} {230604} (\bibinfo {year}
  {2020})}\BibitemShut {NoStop}%
\bibitem [{\citenamefont {Haga}\ \emph {et~al.}(2021)\citenamefont {Haga},
  \citenamefont {Nakagawa}, \citenamefont {Hamazaki},\ and\ \citenamefont
  {Ueda}}]{haga2021liouvillian}%
  \BibitemOpen
  \bibfield  {author} {\bibinfo {author} {\bibfnamefont {T.}~\bibnamefont
  {Haga}}, \bibinfo {author} {\bibfnamefont {M.}~\bibnamefont {Nakagawa}},
  \bibinfo {author} {\bibfnamefont {R.}~\bibnamefont {Hamazaki}},\ and\
  \bibinfo {author} {\bibfnamefont {M.}~\bibnamefont {Ueda}},\ }\bibfield
  {title} {\bibinfo {title} {Liouvillian skin effect: Slowing down of
  relaxation processes without gap closing},\ }\href
  {https://doi.org/10.1103/PhysRevLett.127.070402} {\bibfield  {journal}
  {\bibinfo  {journal} {Phys. Rev. Lett.}\ }\textbf {\bibinfo {volume} {127}},\
  \bibinfo {pages} {070402} (\bibinfo {year} {2021})}\BibitemShut {NoStop}%
\bibitem [{\citenamefont {Holzhey}\ \emph {et~al.}(1994)\citenamefont
  {Holzhey}, \citenamefont {Larsen},\ and\ \citenamefont
  {Wilczek}}]{holzhey1994geometric}%
  \BibitemOpen
  \bibfield  {author} {\bibinfo {author} {\bibfnamefont {C.}~\bibnamefont
  {Holzhey}}, \bibinfo {author} {\bibfnamefont {F.}~\bibnamefont {Larsen}},\
  and\ \bibinfo {author} {\bibfnamefont {F.}~\bibnamefont {Wilczek}},\
  }\bibfield  {title} {\bibinfo {title} {Geometric and renormalized entropy in
  conformal field theory},\ }\href
  {https://doi.org/https://doi.org/10.1016/0550-3213(94)90402-2} {\bibfield
  {journal} {\bibinfo  {journal} {Nuclear physics b}\ }\textbf {\bibinfo
  {volume} {424}},\ \bibinfo {pages} {443} (\bibinfo {year}
  {1994})}\BibitemShut {NoStop}%
\bibitem [{\citenamefont {Calabrese}\ and\ \citenamefont
  {Cardy}(2004)}]{calabrese2004entanglement}%
  \BibitemOpen
  \bibfield  {author} {\bibinfo {author} {\bibfnamefont {P.}~\bibnamefont
  {Calabrese}}\ and\ \bibinfo {author} {\bibfnamefont {J.}~\bibnamefont
  {Cardy}},\ }\bibfield  {title} {\bibinfo {title} {Entanglement entropy and
  quantum field theory},\ }\href
  {https://doi.org/10.1088/1742-5468/2004/06/P06002} {\bibfield  {journal}
  {\bibinfo  {journal} {Journal of statistical mechanics: theory and
  experiment}\ }\textbf {\bibinfo {volume} {2004}},\ \bibinfo {pages} {P06002}
  (\bibinfo {year} {2004})}\BibitemShut {NoStop}%
\bibitem [{\citenamefont {Hastings}\ and\ \citenamefont
  {Koma}(2006)}]{hastings2006spectral}%
  \BibitemOpen
  \bibfield  {author} {\bibinfo {author} {\bibfnamefont {M.~B.}\ \bibnamefont
  {Hastings}}\ and\ \bibinfo {author} {\bibfnamefont {T.}~\bibnamefont
  {Koma}},\ }\bibfield  {title} {\bibinfo {title} {Spectral gap and exponential
  decay of correlations},\ }\href {https://doi.org/10.1007/s00220-006-0030-4}
  {\bibfield  {journal} {\bibinfo  {journal} {Communications in mathematical
  physics}\ }\textbf {\bibinfo {volume} {265}},\ \bibinfo {pages} {781}
  (\bibinfo {year} {2006})}\BibitemShut {NoStop}%
\bibitem [{\citenamefont {Hastings}(2007)}]{hastings2007area}%
  \BibitemOpen
  \bibfield  {author} {\bibinfo {author} {\bibfnamefont {M.~B.}\ \bibnamefont
  {Hastings}},\ }\bibfield  {title} {\bibinfo {title} {An area law for
  one-dimensional quantum systems},\ }\href
  {https://dx.doi.org/10.1088/1742-5468/2007/08/P08024} {\bibfield  {journal}
  {\bibinfo  {journal} {Journal of statistical mechanics: theory and
  experiment}\ }\textbf {\bibinfo {volume} {2007}},\ \bibinfo {pages} {P08024}
  (\bibinfo {year} {2007})}\BibitemShut {NoStop}%
\bibitem [{\citenamefont {Li}\ \emph {et~al.}(2018)\citenamefont {Li},
  \citenamefont {Chen},\ and\ \citenamefont {Fisher}}]{li2018quantum}%
  \BibitemOpen
  \bibfield  {author} {\bibinfo {author} {\bibfnamefont {Y.}~\bibnamefont
  {Li}}, \bibinfo {author} {\bibfnamefont {X.}~\bibnamefont {Chen}},\ and\
  \bibinfo {author} {\bibfnamefont {M.~P.~A.}\ \bibnamefont {Fisher}},\
  }\bibfield  {title} {\bibinfo {title} {Quantum zeno effect and the many-body
  entanglement transition},\ }\href
  {https://doi.org/10.1103/PhysRevB.98.205136} {\bibfield  {journal} {\bibinfo
  {journal} {Phys. Rev. B}\ }\textbf {\bibinfo {volume} {98}},\ \bibinfo
  {pages} {205136} (\bibinfo {year} {2018})}\BibitemShut {NoStop}%
\bibitem [{\citenamefont {Cao}\ \emph {et~al.}(2019)\citenamefont {Cao},
  \citenamefont {Tilloy},\ and\ \citenamefont {Luca}}]{cao2019entanglement}%
  \BibitemOpen
  \bibfield  {author} {\bibinfo {author} {\bibfnamefont {X.}~\bibnamefont
  {Cao}}, \bibinfo {author} {\bibfnamefont {A.}~\bibnamefont {Tilloy}},\ and\
  \bibinfo {author} {\bibfnamefont {A.~D.}\ \bibnamefont {Luca}},\ }\bibfield
  {title} {\bibinfo {title} {{Entanglement in a fermion chain under continuous
  monitoring}},\ }\href {https://doi.org/10.21468/SciPostPhys.7.2.024}
  {\bibfield  {journal} {\bibinfo  {journal} {SciPost Phys.}\ }\textbf
  {\bibinfo {volume} {7}},\ \bibinfo {pages} {024} (\bibinfo {year}
  {2019})}\BibitemShut {NoStop}%
\bibitem [{\citenamefont {Chan}\ \emph {et~al.}(2019)\citenamefont {Chan},
  \citenamefont {Nandkishore}, \citenamefont {Pretko},\ and\ \citenamefont
  {Smith}}]{chan2019unitary}%
  \BibitemOpen
  \bibfield  {author} {\bibinfo {author} {\bibfnamefont {A.}~\bibnamefont
  {Chan}}, \bibinfo {author} {\bibfnamefont {R.~M.}\ \bibnamefont
  {Nandkishore}}, \bibinfo {author} {\bibfnamefont {M.}~\bibnamefont
  {Pretko}},\ and\ \bibinfo {author} {\bibfnamefont {G.}~\bibnamefont
  {Smith}},\ }\bibfield  {title} {\bibinfo {title} {Unitary-projective
  entanglement dynamics},\ }\href {https://doi.org/10.1103/PhysRevB.99.224307}
  {\bibfield  {journal} {\bibinfo  {journal} {Phys. Rev. B}\ }\textbf {\bibinfo
  {volume} {99}},\ \bibinfo {pages} {224307} (\bibinfo {year}
  {2019})}\BibitemShut {NoStop}%
\bibitem [{\citenamefont {Li}\ \emph {et~al.}(2019)\citenamefont {Li},
  \citenamefont {Chen},\ and\ \citenamefont {Fisher}}]{li2019measurement}%
  \BibitemOpen
  \bibfield  {author} {\bibinfo {author} {\bibfnamefont {Y.}~\bibnamefont
  {Li}}, \bibinfo {author} {\bibfnamefont {X.}~\bibnamefont {Chen}},\ and\
  \bibinfo {author} {\bibfnamefont {M.~P.~A.}\ \bibnamefont {Fisher}},\
  }\bibfield  {title} {\bibinfo {title} {Measurement-driven entanglement
  transition in hybrid quantum circuits},\ }\href
  {https://doi.org/10.1103/PhysRevB.100.134306} {\bibfield  {journal} {\bibinfo
   {journal} {Phys. Rev. B}\ }\textbf {\bibinfo {volume} {100}},\ \bibinfo
  {pages} {134306} (\bibinfo {year} {2019})}\BibitemShut {NoStop}%
\bibitem [{\citenamefont {Skinner}\ \emph {et~al.}(2019)\citenamefont
  {Skinner}, \citenamefont {Ruhman},\ and\ \citenamefont
  {Nahum}}]{skinner2019measurement}%
  \BibitemOpen
  \bibfield  {author} {\bibinfo {author} {\bibfnamefont {B.}~\bibnamefont
  {Skinner}}, \bibinfo {author} {\bibfnamefont {J.}~\bibnamefont {Ruhman}},\
  and\ \bibinfo {author} {\bibfnamefont {A.}~\bibnamefont {Nahum}},\ }\bibfield
   {title} {\bibinfo {title} {Measurement-induced phase transitions in the
  dynamics of entanglement},\ }\href
  {https://doi.org/10.1103/PhysRevX.9.031009} {\bibfield  {journal} {\bibinfo
  {journal} {Phys. Rev. X}\ }\textbf {\bibinfo {volume} {9}},\ \bibinfo {pages}
  {031009} (\bibinfo {year} {2019})}\BibitemShut {NoStop}%
\bibitem [{\citenamefont {Szyniszewski}\ \emph {et~al.}(2019)\citenamefont
  {Szyniszewski}, \citenamefont {Romito},\ and\ \citenamefont
  {Schomerus}}]{szyniszewski2019entanglement}%
  \BibitemOpen
  \bibfield  {author} {\bibinfo {author} {\bibfnamefont {M.}~\bibnamefont
  {Szyniszewski}}, \bibinfo {author} {\bibfnamefont {A.}~\bibnamefont
  {Romito}},\ and\ \bibinfo {author} {\bibfnamefont {H.}~\bibnamefont
  {Schomerus}},\ }\bibfield  {title} {\bibinfo {title} {Entanglement transition
  from variable-strength weak measurements},\ }\href
  {https://doi.org/10.1103/PhysRevB.100.064204} {\bibfield  {journal} {\bibinfo
   {journal} {Phys. Rev. B}\ }\textbf {\bibinfo {volume} {100}},\ \bibinfo
  {pages} {064204} (\bibinfo {year} {2019})}\BibitemShut {NoStop}%
\bibitem [{\citenamefont {Bao}\ \emph {et~al.}(2020)\citenamefont {Bao},
  \citenamefont {Choi},\ and\ \citenamefont {Altman}}]{bao2020theory}%
  \BibitemOpen
  \bibfield  {author} {\bibinfo {author} {\bibfnamefont {Y.}~\bibnamefont
  {Bao}}, \bibinfo {author} {\bibfnamefont {S.}~\bibnamefont {Choi}},\ and\
  \bibinfo {author} {\bibfnamefont {E.}~\bibnamefont {Altman}},\ }\bibfield
  {title} {\bibinfo {title} {Theory of the phase transition in random unitary
  circuits with measurements},\ }\href
  {https://doi.org/10.1103/PhysRevB.101.104301} {\bibfield  {journal} {\bibinfo
   {journal} {Phys. Rev. B}\ }\textbf {\bibinfo {volume} {101}},\ \bibinfo
  {pages} {104301} (\bibinfo {year} {2020})}\BibitemShut {NoStop}%
\bibitem [{\citenamefont {Choi}\ \emph {et~al.}(2020)\citenamefont {Choi},
  \citenamefont {Bao}, \citenamefont {Qi},\ and\ \citenamefont
  {Altman}}]{choi2020quantum}%
  \BibitemOpen
  \bibfield  {author} {\bibinfo {author} {\bibfnamefont {S.}~\bibnamefont
  {Choi}}, \bibinfo {author} {\bibfnamefont {Y.}~\bibnamefont {Bao}}, \bibinfo
  {author} {\bibfnamefont {X.-L.}\ \bibnamefont {Qi}},\ and\ \bibinfo {author}
  {\bibfnamefont {E.}~\bibnamefont {Altman}},\ }\bibfield  {title} {\bibinfo
  {title} {Quantum error correction in scrambling dynamics and
  measurement-induced phase transition},\ }\href
  {https://doi.org/10.1103/PhysRevLett.125.030505} {\bibfield  {journal}
  {\bibinfo  {journal} {Phys. Rev. Lett.}\ }\textbf {\bibinfo {volume} {125}},\
  \bibinfo {pages} {030505} (\bibinfo {year} {2020})}\BibitemShut {NoStop}%
\bibitem [{\citenamefont {Fuji}\ and\ \citenamefont
  {Ashida}(2020)}]{fuji2020measurement}%
  \BibitemOpen
  \bibfield  {author} {\bibinfo {author} {\bibfnamefont {Y.}~\bibnamefont
  {Fuji}}\ and\ \bibinfo {author} {\bibfnamefont {Y.}~\bibnamefont {Ashida}},\
  }\bibfield  {title} {\bibinfo {title} {Measurement-induced quantum
  criticality under continuous monitoring},\ }\href
  {https://doi.org/10.1103/PhysRevB.102.054302} {\bibfield  {journal} {\bibinfo
   {journal} {Phys. Rev. B}\ }\textbf {\bibinfo {volume} {102}},\ \bibinfo
  {pages} {054302} (\bibinfo {year} {2020})}\BibitemShut {NoStop}%
\bibitem [{\citenamefont {Gullans}\ and\ \citenamefont
  {Huse}(2020)}]{gullans2020dynamical}%
  \BibitemOpen
  \bibfield  {author} {\bibinfo {author} {\bibfnamefont {M.~J.}\ \bibnamefont
  {Gullans}}\ and\ \bibinfo {author} {\bibfnamefont {D.~A.}\ \bibnamefont
  {Huse}},\ }\bibfield  {title} {\bibinfo {title} {Dynamical purification phase
  transition induced by quantum measurements},\ }\href
  {https://doi.org/10.1103/PhysRevX.10.041020} {\bibfield  {journal} {\bibinfo
  {journal} {Phys. Rev. X}\ }\textbf {\bibinfo {volume} {10}},\ \bibinfo
  {pages} {041020} (\bibinfo {year} {2020})}\BibitemShut {NoStop}%
\bibitem [{\citenamefont {Lunt}\ and\ \citenamefont
  {Pal}(2020)}]{lunt2020measurement}%
  \BibitemOpen
  \bibfield  {author} {\bibinfo {author} {\bibfnamefont {O.}~\bibnamefont
  {Lunt}}\ and\ \bibinfo {author} {\bibfnamefont {A.}~\bibnamefont {Pal}},\
  }\bibfield  {title} {\bibinfo {title} {Measurement-induced entanglement
  transitions in many-body localized systems},\ }\href
  {https://doi.org/10.1103/PhysRevResearch.2.043072} {\bibfield  {journal}
  {\bibinfo  {journal} {Phys. Rev. Res.}\ }\textbf {\bibinfo {volume} {2}},\
  \bibinfo {pages} {043072} (\bibinfo {year} {2020})}\BibitemShut {NoStop}%
\bibitem [{\citenamefont {Szyniszewski}\ \emph {et~al.}(2020)\citenamefont
  {Szyniszewski}, \citenamefont {Romito},\ and\ \citenamefont
  {Schomerus}}]{szyniszewski2020universality}%
  \BibitemOpen
  \bibfield  {author} {\bibinfo {author} {\bibfnamefont {M.}~\bibnamefont
  {Szyniszewski}}, \bibinfo {author} {\bibfnamefont {A.}~\bibnamefont
  {Romito}},\ and\ \bibinfo {author} {\bibfnamefont {H.}~\bibnamefont
  {Schomerus}},\ }\bibfield  {title} {\bibinfo {title} {Universality of
  entanglement transitions from stroboscopic to continuous measurements},\
  }\href {https://doi.org/10.1103/PhysRevLett.125.210602} {\bibfield  {journal}
  {\bibinfo  {journal} {Phys. Rev. Lett.}\ }\textbf {\bibinfo {volume} {125}},\
  \bibinfo {pages} {210602} (\bibinfo {year} {2020})}\BibitemShut {NoStop}%
\bibitem [{\citenamefont {Turkeshi}\ \emph {et~al.}(2020)\citenamefont
  {Turkeshi}, \citenamefont {Fazio},\ and\ \citenamefont
  {Dalmonte}}]{turkeshi2020measurement}%
  \BibitemOpen
  \bibfield  {author} {\bibinfo {author} {\bibfnamefont {X.}~\bibnamefont
  {Turkeshi}}, \bibinfo {author} {\bibfnamefont {R.}~\bibnamefont {Fazio}},\
  and\ \bibinfo {author} {\bibfnamefont {M.}~\bibnamefont {Dalmonte}},\
  }\bibfield  {title} {\bibinfo {title} {Measurement-induced criticality in
  $(2+1)$-dimensional hybrid quantum circuits},\ }\href
  {https://doi.org/10.1103/PhysRevB.102.014315} {\bibfield  {journal} {\bibinfo
   {journal} {Phys. Rev. B}\ }\textbf {\bibinfo {volume} {102}},\ \bibinfo
  {pages} {014315} (\bibinfo {year} {2020})}\BibitemShut {NoStop}%
\bibitem [{\citenamefont {Alberton}\ \emph {et~al.}(2021)\citenamefont
  {Alberton}, \citenamefont {Buchhold},\ and\ \citenamefont
  {Diehl}}]{alberton2021entanglement}%
  \BibitemOpen
  \bibfield  {author} {\bibinfo {author} {\bibfnamefont {O.}~\bibnamefont
  {Alberton}}, \bibinfo {author} {\bibfnamefont {M.}~\bibnamefont {Buchhold}},\
  and\ \bibinfo {author} {\bibfnamefont {S.}~\bibnamefont {Diehl}},\ }\bibfield
   {title} {\bibinfo {title} {Entanglement transition in a monitored
  free-fermion chain: From extended criticality to area law},\ }\href
  {https://doi.org/10.1103/PhysRevLett.126.170602} {\bibfield  {journal}
  {\bibinfo  {journal} {Phys. Rev. Lett.}\ }\textbf {\bibinfo {volume} {126}},\
  \bibinfo {pages} {170602} (\bibinfo {year} {2021})}\BibitemShut {NoStop}%
\bibitem [{\citenamefont {Lu}\ and\ \citenamefont
  {Grover}(2021)}]{lu2021spacetime}%
  \BibitemOpen
  \bibfield  {author} {\bibinfo {author} {\bibfnamefont {T.-C.}\ \bibnamefont
  {Lu}}\ and\ \bibinfo {author} {\bibfnamefont {T.}~\bibnamefont {Grover}},\
  }\bibfield  {title} {\bibinfo {title} {Spacetime duality between localization
  transitions and measurement-induced transitions},\ }\href
  {https://doi.org/10.1103/PRXQuantum.2.040319} {\bibfield  {journal} {\bibinfo
   {journal} {PRX Quantum}\ }\textbf {\bibinfo {volume} {2}},\ \bibinfo {pages}
  {040319} (\bibinfo {year} {2021})}\BibitemShut {NoStop}%
\bibitem [{\citenamefont {Agrawal}\ \emph {et~al.}(2022)\citenamefont
  {Agrawal}, \citenamefont {Zabalo}, \citenamefont {Chen}, \citenamefont
  {Wilson}, \citenamefont {Potter}, \citenamefont {Pixley}, \citenamefont
  {Gopalakrishnan},\ and\ \citenamefont {Vasseur}}]{agrawal2022entanglement}%
  \BibitemOpen
  \bibfield  {author} {\bibinfo {author} {\bibfnamefont {U.}~\bibnamefont
  {Agrawal}}, \bibinfo {author} {\bibfnamefont {A.}~\bibnamefont {Zabalo}},
  \bibinfo {author} {\bibfnamefont {K.}~\bibnamefont {Chen}}, \bibinfo {author}
  {\bibfnamefont {J.~H.}\ \bibnamefont {Wilson}}, \bibinfo {author}
  {\bibfnamefont {A.~C.}\ \bibnamefont {Potter}}, \bibinfo {author}
  {\bibfnamefont {J.~H.}\ \bibnamefont {Pixley}}, \bibinfo {author}
  {\bibfnamefont {S.}~\bibnamefont {Gopalakrishnan}},\ and\ \bibinfo {author}
  {\bibfnamefont {R.}~\bibnamefont {Vasseur}},\ }\bibfield  {title} {\bibinfo
  {title} {Entanglement and charge-sharpening transitions in u(1) symmetric
  monitored quantum circuits},\ }\href
  {https://doi.org/10.1103/PhysRevX.12.041002} {\bibfield  {journal} {\bibinfo
  {journal} {Phys. Rev. X}\ }\textbf {\bibinfo {volume} {12}},\ \bibinfo
  {pages} {041002} (\bibinfo {year} {2022})}\BibitemShut {NoStop}%
\bibitem [{\citenamefont {Barratt}\ \emph {et~al.}(2022)\citenamefont
  {Barratt}, \citenamefont {Agrawal}, \citenamefont {Gopalakrishnan},
  \citenamefont {Huse}, \citenamefont {Vasseur},\ and\ \citenamefont
  {Potter}}]{barratt2022field}%
  \BibitemOpen
  \bibfield  {author} {\bibinfo {author} {\bibfnamefont {F.}~\bibnamefont
  {Barratt}}, \bibinfo {author} {\bibfnamefont {U.}~\bibnamefont {Agrawal}},
  \bibinfo {author} {\bibfnamefont {S.}~\bibnamefont {Gopalakrishnan}},
  \bibinfo {author} {\bibfnamefont {D.~A.}\ \bibnamefont {Huse}}, \bibinfo
  {author} {\bibfnamefont {R.}~\bibnamefont {Vasseur}},\ and\ \bibinfo {author}
  {\bibfnamefont {A.~C.}\ \bibnamefont {Potter}},\ }\bibfield  {title}
  {\bibinfo {title} {Field theory of charge sharpening in symmetric monitored
  quantum circuits},\ }\href {https://doi.org/10.1103/PhysRevLett.129.120604}
  {\bibfield  {journal} {\bibinfo  {journal} {Phys. Rev. Lett.}\ }\textbf
  {\bibinfo {volume} {129}},\ \bibinfo {pages} {120604} (\bibinfo {year}
  {2022})}\BibitemShut {NoStop}%
\bibitem [{\citenamefont {Block}\ \emph {et~al.}(2022)\citenamefont {Block},
  \citenamefont {Bao}, \citenamefont {Choi}, \citenamefont {Altman},\ and\
  \citenamefont {Yao}}]{block2022measurement}%
  \BibitemOpen
  \bibfield  {author} {\bibinfo {author} {\bibfnamefont {M.}~\bibnamefont
  {Block}}, \bibinfo {author} {\bibfnamefont {Y.}~\bibnamefont {Bao}}, \bibinfo
  {author} {\bibfnamefont {S.}~\bibnamefont {Choi}}, \bibinfo {author}
  {\bibfnamefont {E.}~\bibnamefont {Altman}},\ and\ \bibinfo {author}
  {\bibfnamefont {N.~Y.}\ \bibnamefont {Yao}},\ }\bibfield  {title} {\bibinfo
  {title} {Measurement-induced transition in long-range interacting quantum
  circuits},\ }\href {https://doi.org/10.1103/PhysRevLett.128.010604}
  {\bibfield  {journal} {\bibinfo  {journal} {Phys. Rev. Lett.}\ }\textbf
  {\bibinfo {volume} {128}},\ \bibinfo {pages} {010604} (\bibinfo {year}
  {2022})}\BibitemShut {NoStop}%
\bibitem [{\citenamefont {Minato}\ \emph {et~al.}(2022)\citenamefont {Minato},
  \citenamefont {Sugimoto}, \citenamefont {Kuwahara},\ and\ \citenamefont
  {Saito}}]{minato2022fate}%
  \BibitemOpen
  \bibfield  {author} {\bibinfo {author} {\bibfnamefont {T.}~\bibnamefont
  {Minato}}, \bibinfo {author} {\bibfnamefont {K.}~\bibnamefont {Sugimoto}},
  \bibinfo {author} {\bibfnamefont {T.}~\bibnamefont {Kuwahara}},\ and\
  \bibinfo {author} {\bibfnamefont {K.}~\bibnamefont {Saito}},\ }\bibfield
  {title} {\bibinfo {title} {Fate of measurement-induced phase transition in
  long-range interactions},\ }\href
  {https://doi.org/10.1103/PhysRevLett.128.010603} {\bibfield  {journal}
  {\bibinfo  {journal} {Phys. Rev. Lett.}\ }\textbf {\bibinfo {volume} {128}},\
  \bibinfo {pages} {010603} (\bibinfo {year} {2022})}\BibitemShut {NoStop}%
\bibitem [{\citenamefont {M\"uller}\ \emph {et~al.}(2022)\citenamefont
  {M\"uller}, \citenamefont {Diehl},\ and\ \citenamefont
  {Buchhold}}]{muller2022measurement}%
  \BibitemOpen
  \bibfield  {author} {\bibinfo {author} {\bibfnamefont {T.}~\bibnamefont
  {M\"uller}}, \bibinfo {author} {\bibfnamefont {S.}~\bibnamefont {Diehl}},\
  and\ \bibinfo {author} {\bibfnamefont {M.}~\bibnamefont {Buchhold}},\
  }\bibfield  {title} {\bibinfo {title} {Measurement-induced dark state phase
  transitions in long-ranged fermion systems},\ }\href
  {https://doi.org/10.1103/PhysRevLett.128.010605} {\bibfield  {journal}
  {\bibinfo  {journal} {Phys. Rev. Lett.}\ }\textbf {\bibinfo {volume} {128}},\
  \bibinfo {pages} {010605} (\bibinfo {year} {2022})}\BibitemShut {NoStop}%
\bibitem [{\citenamefont {Noel}\ \emph {et~al.}(2022)\citenamefont {Noel},
  \citenamefont {Niroula}, \citenamefont {Zhu}, \citenamefont {Risinger},
  \citenamefont {Egan}, \citenamefont {Biswas}, \citenamefont {Cetina},
  \citenamefont {Gorshkov}, \citenamefont {Gullans}, \citenamefont {Huse},\
  and\ \citenamefont {Monroe}}]{noel2022measurement}%
  \BibitemOpen
  \bibfield  {author} {\bibinfo {author} {\bibfnamefont {C.}~\bibnamefont
  {Noel}}, \bibinfo {author} {\bibfnamefont {P.}~\bibnamefont {Niroula}},
  \bibinfo {author} {\bibfnamefont {D.}~\bibnamefont {Zhu}}, \bibinfo {author}
  {\bibfnamefont {A.}~\bibnamefont {Risinger}}, \bibinfo {author}
  {\bibfnamefont {L.}~\bibnamefont {Egan}}, \bibinfo {author} {\bibfnamefont
  {D.}~\bibnamefont {Biswas}}, \bibinfo {author} {\bibfnamefont
  {M.}~\bibnamefont {Cetina}}, \bibinfo {author} {\bibfnamefont {A.~V.}\
  \bibnamefont {Gorshkov}}, \bibinfo {author} {\bibfnamefont {M.~J.}\
  \bibnamefont {Gullans}}, \bibinfo {author} {\bibfnamefont {D.~A.}\
  \bibnamefont {Huse}},\ and\ \bibinfo {author} {\bibfnamefont
  {C.}~\bibnamefont {Monroe}},\ }\bibfield  {title} {\bibinfo {title}
  {Measurement-induced quantum phases realized in a trapped-ion quantum
  computer},\ }\href {https://doi.org/10.1038/s41567-022-01619-7} {\bibfield
  {journal} {\bibinfo  {journal} {Nature Physics}\ }\textbf {\bibinfo {volume}
  {18}},\ \bibinfo {pages} {760} (\bibinfo {year} {2022})}\BibitemShut
  {NoStop}%
\bibitem [{\citenamefont {Sierant}\ and\ \citenamefont
  {Turkeshi}(2022)}]{sierant2022universal}%
  \BibitemOpen
  \bibfield  {author} {\bibinfo {author} {\bibfnamefont {P.}~\bibnamefont
  {Sierant}}\ and\ \bibinfo {author} {\bibfnamefont {X.}~\bibnamefont
  {Turkeshi}},\ }\bibfield  {title} {\bibinfo {title} {Universal behavior
  beyond multifractality of wave functions at measurement-induced phase
  transitions},\ }\href {https://doi.org/10.1103/PhysRevLett.128.130605}
  {\bibfield  {journal} {\bibinfo  {journal} {Phys. Rev. Lett.}\ }\textbf
  {\bibinfo {volume} {128}},\ \bibinfo {pages} {130605} (\bibinfo {year}
  {2022})}\BibitemShut {NoStop}%
\bibitem [{\citenamefont {Zabalo}\ \emph {et~al.}(2022)\citenamefont {Zabalo},
  \citenamefont {Gullans}, \citenamefont {Wilson}, \citenamefont {Vasseur},
  \citenamefont {Ludwig}, \citenamefont {Gopalakrishnan}, \citenamefont
  {Huse},\ and\ \citenamefont {Pixley}}]{zabalo2022operator}%
  \BibitemOpen
  \bibfield  {author} {\bibinfo {author} {\bibfnamefont {A.}~\bibnamefont
  {Zabalo}}, \bibinfo {author} {\bibfnamefont {M.~J.}\ \bibnamefont {Gullans}},
  \bibinfo {author} {\bibfnamefont {J.~H.}\ \bibnamefont {Wilson}}, \bibinfo
  {author} {\bibfnamefont {R.}~\bibnamefont {Vasseur}}, \bibinfo {author}
  {\bibfnamefont {A.~W.~W.}\ \bibnamefont {Ludwig}}, \bibinfo {author}
  {\bibfnamefont {S.}~\bibnamefont {Gopalakrishnan}}, \bibinfo {author}
  {\bibfnamefont {D.~A.}\ \bibnamefont {Huse}},\ and\ \bibinfo {author}
  {\bibfnamefont {J.~H.}\ \bibnamefont {Pixley}},\ }\bibfield  {title}
  {\bibinfo {title} {Operator scaling dimensions and multifractality at
  measurement-induced transitions},\ }\href
  {https://doi.org/10.1103/PhysRevLett.128.050602} {\bibfield  {journal}
  {\bibinfo  {journal} {Phys. Rev. Lett.}\ }\textbf {\bibinfo {volume} {128}},\
  \bibinfo {pages} {050602} (\bibinfo {year} {2022})}\BibitemShut {NoStop}%
\bibitem [{\citenamefont {Granet}\ \emph {et~al.}(2023)\citenamefont {Granet},
  \citenamefont {Zhang},\ and\ \citenamefont {Dreyer}}]{granet2023volume}%
  \BibitemOpen
  \bibfield  {author} {\bibinfo {author} {\bibfnamefont {E.}~\bibnamefont
  {Granet}}, \bibinfo {author} {\bibfnamefont {C.}~\bibnamefont {Zhang}},\ and\
  \bibinfo {author} {\bibfnamefont {H.}~\bibnamefont {Dreyer}},\ }\bibfield
  {title} {\bibinfo {title} {Volume-law to area-law entanglement transition in
  a nonunitary periodic gaussian circuit},\ }\href
  {https://doi.org/10.1103/PhysRevLett.130.230401} {\bibfield  {journal}
  {\bibinfo  {journal} {Phys. Rev. Lett.}\ }\textbf {\bibinfo {volume} {130}},\
  \bibinfo {pages} {230401} (\bibinfo {year} {2023})}\BibitemShut {NoStop}%
\bibitem [{\citenamefont {Koh}\ \emph {et~al.}(2023)\citenamefont {Koh},
  \citenamefont {Sun}, \citenamefont {Motta},\ and\ \citenamefont
  {Minnich}}]{koh2023measurement}%
  \BibitemOpen
  \bibfield  {author} {\bibinfo {author} {\bibfnamefont {J.~M.}\ \bibnamefont
  {Koh}}, \bibinfo {author} {\bibfnamefont {S.-N.}\ \bibnamefont {Sun}},
  \bibinfo {author} {\bibfnamefont {M.}~\bibnamefont {Motta}},\ and\ \bibinfo
  {author} {\bibfnamefont {A.~J.}\ \bibnamefont {Minnich}},\ }\bibfield
  {title} {\bibinfo {title} {Measurement-induced entanglement phase transition
  on a superconducting quantum processor with mid-circuit readout},\ }\href
  {https://doi.org/10.1038/s41567-023-02076-6} {\bibfield  {journal} {\bibinfo
  {journal} {Nature Physics}\ }\textbf {\bibinfo {volume} {19}},\ \bibinfo
  {pages} {1314} (\bibinfo {year} {2023})}\BibitemShut {NoStop}%
\bibitem [{\citenamefont {Le~Gal}\ \emph {et~al.}(2023)\citenamefont {Le~Gal},
  \citenamefont {Turkeshi},\ and\ \citenamefont {Schir{\`o}}}]{le2023volume}%
  \BibitemOpen
  \bibfield  {author} {\bibinfo {author} {\bibfnamefont {Y.}~\bibnamefont
  {Le~Gal}}, \bibinfo {author} {\bibfnamefont {X.}~\bibnamefont {Turkeshi}},\
  and\ \bibinfo {author} {\bibfnamefont {M.}~\bibnamefont {Schir{\`o}}},\
  }\bibfield  {title} {\bibinfo {title} {Volume-to-area law entanglement
  transition in a non-{H}ermitian free fermionic chain},\ }\href
  {https://scipost.org/10.21468/SciPostPhys.14.5.138} {\bibfield  {journal}
  {\bibinfo  {journal} {SciPost Physics}\ }\textbf {\bibinfo {volume} {14}},\
  \bibinfo {pages} {138} (\bibinfo {year} {2023})}\BibitemShut {NoStop}%
\bibitem [{\citenamefont {L\'oio}\ \emph {et~al.}(2023)\citenamefont {L\'oio},
  \citenamefont {De~Luca}, \citenamefont {De~Nardis},\ and\ \citenamefont
  {Turkeshi}}]{loio2023purification}%
  \BibitemOpen
  \bibfield  {author} {\bibinfo {author} {\bibfnamefont {H.}~\bibnamefont
  {L\'oio}}, \bibinfo {author} {\bibfnamefont {A.}~\bibnamefont {De~Luca}},
  \bibinfo {author} {\bibfnamefont {J.}~\bibnamefont {De~Nardis}},\ and\
  \bibinfo {author} {\bibfnamefont {X.}~\bibnamefont {Turkeshi}},\ }\bibfield
  {title} {\bibinfo {title} {Purification timescales in monitored fermions},\
  }\href {https://doi.org/10.1103/PhysRevB.108.L020306} {\bibfield  {journal}
  {\bibinfo  {journal} {Phys. Rev. B}\ }\textbf {\bibinfo {volume} {108}},\
  \bibinfo {pages} {L020306} (\bibinfo {year} {2023})}\BibitemShut {NoStop}%
\bibitem [{\citenamefont {Majidy}\ \emph {et~al.}(2023)\citenamefont {Majidy},
  \citenamefont {Agrawal}, \citenamefont {Gopalakrishnan}, \citenamefont
  {Potter}, \citenamefont {Vasseur},\ and\ \citenamefont
  {Halpern}}]{majidy2023critical}%
  \BibitemOpen
  \bibfield  {author} {\bibinfo {author} {\bibfnamefont {S.}~\bibnamefont
  {Majidy}}, \bibinfo {author} {\bibfnamefont {U.}~\bibnamefont {Agrawal}},
  \bibinfo {author} {\bibfnamefont {S.}~\bibnamefont {Gopalakrishnan}},
  \bibinfo {author} {\bibfnamefont {A.~C.}\ \bibnamefont {Potter}}, \bibinfo
  {author} {\bibfnamefont {R.}~\bibnamefont {Vasseur}},\ and\ \bibinfo {author}
  {\bibfnamefont {N.~Y.}\ \bibnamefont {Halpern}},\ }\bibfield  {title}
  {\bibinfo {title} {Critical phase and spin sharpening in su(2)-symmetric
  monitored quantum circuits},\ }\href
  {https://doi.org/10.1103/PhysRevB.108.054307} {\bibfield  {journal} {\bibinfo
   {journal} {Phys. Rev. B}\ }\textbf {\bibinfo {volume} {108}},\ \bibinfo
  {pages} {054307} (\bibinfo {year} {2023})}\BibitemShut {NoStop}%
\bibitem [{\citenamefont {Oshima}\ and\ \citenamefont
  {Fuji}(2023)}]{oshima2023charge}%
  \BibitemOpen
  \bibfield  {author} {\bibinfo {author} {\bibfnamefont {H.}~\bibnamefont
  {Oshima}}\ and\ \bibinfo {author} {\bibfnamefont {Y.}~\bibnamefont {Fuji}},\
  }\bibfield  {title} {\bibinfo {title} {Charge fluctuation and charge-resolved
  entanglement in a monitored quantum circuit with $u(1)$ symmetry},\ }\href
  {https://doi.org/10.1103/PhysRevB.107.014308} {\bibfield  {journal} {\bibinfo
   {journal} {Phys. Rev. B}\ }\textbf {\bibinfo {volume} {107}},\ \bibinfo
  {pages} {014308} (\bibinfo {year} {2023})}\BibitemShut {NoStop}%
\bibitem [{\citenamefont {Yamamoto}\ and\ \citenamefont
  {Hamazaki}(2023)}]{yamamoto2023localization}%
  \BibitemOpen
  \bibfield  {author} {\bibinfo {author} {\bibfnamefont {K.}~\bibnamefont
  {Yamamoto}}\ and\ \bibinfo {author} {\bibfnamefont {R.}~\bibnamefont
  {Hamazaki}},\ }\bibfield  {title} {\bibinfo {title} {Localization properties
  in disordered quantum many-body dynamics under continuous measurement},\
  }\href {https://doi.org/10.1103/PhysRevB.107.L220201} {\bibfield  {journal}
  {\bibinfo  {journal} {Phys. Rev. B}\ }\textbf {\bibinfo {volume} {107}},\
  \bibinfo {pages} {L220201} (\bibinfo {year} {2023})}\BibitemShut {NoStop}%
\bibitem [{\citenamefont {Kumar}\ \emph {et~al.}(2024)\citenamefont {Kumar},
  \citenamefont {Aziz}, \citenamefont {Chakraborty}, \citenamefont {Ludwig},
  \citenamefont {Gopalakrishnan}, \citenamefont {Pixley},\ and\ \citenamefont
  {Vasseur}}]{kumar2024boundary}%
  \BibitemOpen
  \bibfield  {author} {\bibinfo {author} {\bibfnamefont {A.}~\bibnamefont
  {Kumar}}, \bibinfo {author} {\bibfnamefont {K.}~\bibnamefont {Aziz}},
  \bibinfo {author} {\bibfnamefont {A.}~\bibnamefont {Chakraborty}}, \bibinfo
  {author} {\bibfnamefont {A.~W.~W.}\ \bibnamefont {Ludwig}}, \bibinfo {author}
  {\bibfnamefont {S.}~\bibnamefont {Gopalakrishnan}}, \bibinfo {author}
  {\bibfnamefont {J.~H.}\ \bibnamefont {Pixley}},\ and\ \bibinfo {author}
  {\bibfnamefont {R.}~\bibnamefont {Vasseur}},\ }\bibfield  {title} {\bibinfo
  {title} {Boundary transfer matrix spectrum of measurement-induced
  transitions},\ }\href {https://doi.org/10.1103/PhysRevB.109.014303}
  {\bibfield  {journal} {\bibinfo  {journal} {Phys. Rev. B}\ }\textbf {\bibinfo
  {volume} {109}},\ \bibinfo {pages} {014303} (\bibinfo {year}
  {2024})}\BibitemShut {NoStop}%
\bibitem [{\citenamefont {Aziz}\ \emph {et~al.}(2024)\citenamefont {Aziz},
  \citenamefont {Chakraborty},\ and\ \citenamefont
  {Pixley}}]{aziz2024critical}%
  \BibitemOpen
  \bibfield  {author} {\bibinfo {author} {\bibfnamefont {K.}~\bibnamefont
  {Aziz}}, \bibinfo {author} {\bibfnamefont {A.}~\bibnamefont {Chakraborty}},\
  and\ \bibinfo {author} {\bibfnamefont {J.~H.}\ \bibnamefont {Pixley}},\
  }\bibfield  {title} {\bibinfo {title} {Critical properties of weak
  measurement induced phase transitions in random quantum circuits},\ }\href
  {https://doi.org/10.1103/PhysRevB.110.064301} {\bibfield  {journal} {\bibinfo
   {journal} {Phys. Rev. B}\ }\textbf {\bibinfo {volume} {110}},\ \bibinfo
  {pages} {064301} (\bibinfo {year} {2024})}\BibitemShut {NoStop}%
\bibitem [{\citenamefont {Mochizuki}\ and\ \citenamefont
  {Hamazaki}(2023)}]{mochizuki2023distinguishability}%
  \BibitemOpen
  \bibfield  {author} {\bibinfo {author} {\bibfnamefont {K.}~\bibnamefont
  {Mochizuki}}\ and\ \bibinfo {author} {\bibfnamefont {R.}~\bibnamefont
  {Hamazaki}},\ }\bibfield  {title} {\bibinfo {title} {Distinguishability
  transitions in nonunitary boson-sampling dynamics},\ }\href
  {https://doi.org/10.1103/PhysRevResearch.5.013177} {\bibfield  {journal}
  {\bibinfo  {journal} {Phys. Rev. Res.}\ }\textbf {\bibinfo {volume} {5}},\
  \bibinfo {pages} {013177} (\bibinfo {year} {2023})}\BibitemShut {NoStop}%
\bibitem
  [{sup()}]{supplemental-material_measurement-induced_spectral-transition}%
  \BibitemOpen
  \href@noop {} {\bibinfo  {journal} {See Supplemental Material for how to
  realize the generarized measurement, numerics of monitored dynamics,
  computation of the Lyapunov exponents, estimation of the spectral gap in the
  thermodynamic limit, derivation of the spectral gap when unitary gates are
  absent, and derivation of the upper bound for the width of the Lyapunov
  spectrum, which includes Refs.
  \cite{crisanti2012products,mochizuki2024absorption,hiai2024log}}\
  }\BibitemShut {NoStop}%
\bibitem [{\citenamefont {Ershov}\ and\ \citenamefont
  {Potapov}(1998)}]{ershov1998concept}%
  \BibitemOpen
\bibfield  {journal} {  }\bibfield  {author} {\bibinfo {author} {\bibfnamefont
  {S.~V.}\ \bibnamefont {Ershov}}\ and\ \bibinfo {author} {\bibfnamefont
  {A.~B.}\ \bibnamefont {Potapov}},\ }\bibfield  {title} {\bibinfo {title} {On
  the concept of stationary lyapunov basis},\ }\href
  {https://doi.org/https://doi.org/10.1016/S0167-2789(98)00013-X} {\bibfield
  {journal} {\bibinfo  {journal} {Physica D: Nonlinear Phenomena}\ }\textbf
  {\bibinfo {volume} {118}},\ \bibinfo {pages} {167} (\bibinfo {year}
  {1998})}\BibitemShut {NoStop}%
\bibitem [{\citenamefont {De~Luca}\ \emph {et~al.}(2023)\citenamefont
  {De~Luca}, \citenamefont {Liu}, \citenamefont {Nahum},\ and\ \citenamefont
  {Zhou}}]{de2023universality}%
  \BibitemOpen
  \bibfield  {author} {\bibinfo {author} {\bibfnamefont {A.}~\bibnamefont
  {De~Luca}}, \bibinfo {author} {\bibfnamefont {C.}~\bibnamefont {Liu}},
  \bibinfo {author} {\bibfnamefont {A.}~\bibnamefont {Nahum}},\ and\ \bibinfo
  {author} {\bibfnamefont {T.}~\bibnamefont {Zhou}},\ }\bibfield  {title}
  {\bibinfo {title} {Universality classes for purification in nonunitary
  quantum processes},\ }\href@noop {} {\bibfield  {journal} {\bibinfo
  {journal} {arXiv preprint arXiv:2312.17744}\ } (\bibinfo {year}
  {2023})}\BibitemShut {NoStop}%
\bibitem [{\citenamefont {Bulchandani}\ \emph {et~al.}(2024)\citenamefont
  {Bulchandani}, \citenamefont {Sondhi},\ and\ \citenamefont
  {Chalker}}]{bulchandani2024random}%
  \BibitemOpen
  \bibfield  {author} {\bibinfo {author} {\bibfnamefont {V.~B.}\ \bibnamefont
  {Bulchandani}}, \bibinfo {author} {\bibfnamefont {S.}~\bibnamefont
  {Sondhi}},\ and\ \bibinfo {author} {\bibfnamefont {J.}~\bibnamefont
  {Chalker}},\ }\bibfield  {title} {\bibinfo {title} {Random-matrix models of
  monitored quantum circuits},\ }\href
  {https://doi.org/10.1007/s10955-024-03273-0} {\bibfield  {journal} {\bibinfo
  {journal} {Journal of Statistical Physics}\ }\textbf {\bibinfo {volume}
  {191}},\ \bibinfo {pages} {55} (\bibinfo {year} {2024})}\BibitemShut
  {NoStop}%
\bibitem [{\citenamefont {Zeng}\ \emph {et~al.}(2019)\citenamefont {Zeng},
  \citenamefont {Chen}, \citenamefont {Zhou},\ and\ \citenamefont
  {Wen}}]{zeng2019quantum}%
  \BibitemOpen
  \bibfield  {author} {\bibinfo {author} {\bibfnamefont {B.}~\bibnamefont
  {Zeng}}, \bibinfo {author} {\bibfnamefont {X.}~\bibnamefont {Chen}}, \bibinfo
  {author} {\bibfnamefont {D.-L.}\ \bibnamefont {Zhou}},\ and\ \bibinfo
  {author} {\bibfnamefont {X.-G.}\ \bibnamefont {Wen}},\ }\href@noop {} {\emph
  {\bibinfo {title} {Quantum information meets quantum matter}}}\ (\bibinfo
  {publisher} {Springer},\ \bibinfo {year} {2019})\BibitemShut {NoStop}%
\bibitem [{\citenamefont {Wolf}\ \emph {et~al.}(2008)\citenamefont {Wolf},
  \citenamefont {Verstraete}, \citenamefont {Hastings},\ and\ \citenamefont
  {Cirac}}]{walf2008area}%
  \BibitemOpen
  \bibfield  {author} {\bibinfo {author} {\bibfnamefont {M.~M.}\ \bibnamefont
  {Wolf}}, \bibinfo {author} {\bibfnamefont {F.}~\bibnamefont {Verstraete}},
  \bibinfo {author} {\bibfnamefont {M.~B.}\ \bibnamefont {Hastings}},\ and\
  \bibinfo {author} {\bibfnamefont {J.~I.}\ \bibnamefont {Cirac}},\ }\bibfield
  {title} {\bibinfo {title} {Area laws in quantum systems: Mutual information
  and correlations},\ }\href {https://doi.org/10.1103/PhysRevLett.100.070502}
  {\bibfield  {journal} {\bibinfo  {journal} {Phys. Rev. Lett.}\ }\textbf
  {\bibinfo {volume} {100}},\ \bibinfo {pages} {070502} (\bibinfo {year}
  {2008})}\BibitemShut {NoStop}%
\bibitem [{\citenamefont {Eisert}\ \emph {et~al.}(2010)\citenamefont {Eisert},
  \citenamefont {Cramer},\ and\ \citenamefont {Plenio}}]{eisert2010colloquium}%
  \BibitemOpen
  \bibfield  {author} {\bibinfo {author} {\bibfnamefont {J.}~\bibnamefont
  {Eisert}}, \bibinfo {author} {\bibfnamefont {M.}~\bibnamefont {Cramer}},\
  and\ \bibinfo {author} {\bibfnamefont {M.~B.}\ \bibnamefont {Plenio}},\
  }\bibfield  {title} {\bibinfo {title} {Colloquium: Area laws for the
  entanglement entropy},\ }\href {https://doi.org/10.1103/RevModPhys.82.277}
  {\bibfield  {journal} {\bibinfo  {journal} {Rev. Mod. Phys.}\ }\textbf
  {\bibinfo {volume} {82}},\ \bibinfo {pages} {277} (\bibinfo {year}
  {2010})}\BibitemShut {NoStop}%
\bibitem [{\citenamefont {Hiai}(2024)}]{hiai2024log}%
  \BibitemOpen
  \bibfield  {author} {\bibinfo {author} {\bibfnamefont {F.}~\bibnamefont
  {Hiai}},\ }\bibfield  {title} {\bibinfo {title} {Log-majorization and matrix
  norm inequalities with application to quantum information},\ }\href
  {https://doi.org/10.1007/s44146-024-00142-w} {\bibfield  {journal} {\bibinfo
  {journal} {Acta Scientiarum Mathematicarum}\ ,\ \bibinfo {pages} {(Szeged)}}
  (\bibinfo {year} {2024})}\BibitemShut {NoStop}%
\bibitem [{\citenamefont {Gottesman}\ and\ \citenamefont
  {Hastings}(2010)}]{gottesman2010entanglement}%
  \BibitemOpen
  \bibfield  {author} {\bibinfo {author} {\bibfnamefont {D.}~\bibnamefont
  {Gottesman}}\ and\ \bibinfo {author} {\bibfnamefont {M.~B.}\ \bibnamefont
  {Hastings}},\ }\bibfield  {title} {\bibinfo {title} {Entanglement versus gap
  for one-dimensional spin systems},\ }\href
  {https://dx.doi.org/10.1088/1367-2630/12/2/025002} {\bibfield  {journal}
  {\bibinfo  {journal} {New journal of physics}\ }\textbf {\bibinfo {volume}
  {12}},\ \bibinfo {pages} {025002} (\bibinfo {year} {2010})}\BibitemShut
  {NoStop}%
\bibitem [{\citenamefont {Irani}(2010)}]{irani2010ground}%
  \BibitemOpen
  \bibfield  {author} {\bibinfo {author} {\bibfnamefont {S.}~\bibnamefont
  {Irani}},\ }\bibfield  {title} {\bibinfo {title} {Ground state entanglement
  in one-dimensional translationally invariant quantum systems},\ }\href
  {https://doi.org/10.1063/1.3254321} {\bibfield  {journal} {\bibinfo
  {journal} {Journal of Mathematical Physics}\ }\textbf {\bibinfo {volume}
  {51}} (\bibinfo {year} {2010})}\BibitemShut {NoStop}%
\bibitem [{\citenamefont {Vitagliano}\ \emph {et~al.}(2010)\citenamefont
  {Vitagliano}, \citenamefont {Riera},\ and\ \citenamefont
  {Latorre}}]{vitagliano2010volume}%
  \BibitemOpen
  \bibfield  {author} {\bibinfo {author} {\bibfnamefont {G.}~\bibnamefont
  {Vitagliano}}, \bibinfo {author} {\bibfnamefont {A.}~\bibnamefont {Riera}},\
  and\ \bibinfo {author} {\bibfnamefont {J.~I.}\ \bibnamefont {Latorre}},\
  }\bibfield  {title} {\bibinfo {title} {Volume-law scaling for the
  entanglement entropy in spin-1/2 chains},\ }\href
  {https://dx.doi.org/10.1088/1367-2630/12/11/113049} {\bibfield  {journal}
  {\bibinfo  {journal} {New Journal of Physics}\ }\textbf {\bibinfo {volume}
  {12}},\ \bibinfo {pages} {113049} (\bibinfo {year} {2010})}\BibitemShut
  {NoStop}%
\bibitem [{\citenamefont {Ram{\'\i}rez}\ \emph {et~al.}(2014)\citenamefont
  {Ram{\'\i}rez}, \citenamefont {Rodr{\'\i}guez-Laguna},\ and\ \citenamefont
  {Sierra}}]{ramirez2014conformal}%
  \BibitemOpen
  \bibfield  {author} {\bibinfo {author} {\bibfnamefont {G.}~\bibnamefont
  {Ram{\'\i}rez}}, \bibinfo {author} {\bibfnamefont {J.}~\bibnamefont
  {Rodr{\'\i}guez-Laguna}},\ and\ \bibinfo {author} {\bibfnamefont
  {G.}~\bibnamefont {Sierra}},\ }\bibfield  {title} {\bibinfo {title} {From
  conformal to volume law for the entanglement entropy in exponentially
  deformed critical spin 1/2 chains},\ }\href
  {https://dx.doi.org/10.1088/1742-5468/2014/10/P10004} {\bibfield  {journal}
  {\bibinfo  {journal} {Journal of Statistical Mechanics: Theory and
  Experiment}\ }\textbf {\bibinfo {volume} {2014}},\ \bibinfo {pages} {P10004}
  (\bibinfo {year} {2014})}\BibitemShut {NoStop}%
\bibitem [{\citenamefont {Udagawa}\ and\ \citenamefont
  {Katsura}(2017)}]{udagawa2017finite}%
  \BibitemOpen
  \bibfield  {author} {\bibinfo {author} {\bibfnamefont {T.}~\bibnamefont
  {Udagawa}}\ and\ \bibinfo {author} {\bibfnamefont {H.}~\bibnamefont
  {Katsura}},\ }\bibfield  {title} {\bibinfo {title} {Finite-size gap,
  magnetization, and entanglement of deformed fredkin spin chain},\ }\href
  {https://doi.org/https://iopscience.iop.org/article/10.1088/1751-8121/aa85b5/meta}
  {\bibfield  {journal} {\bibinfo  {journal} {Journal of Physics A:
  Mathematical and Theoretical}\ }\textbf {\bibinfo {volume} {50}},\ \bibinfo
  {pages} {405002} (\bibinfo {year} {2017})}\BibitemShut {NoStop}%
\bibitem [{\citenamefont {Zhang}\ \emph {et~al.}(2017)\citenamefont {Zhang},
  \citenamefont {Ahmadain},\ and\ \citenamefont {Klich}}]{zhang2017novel}%
  \BibitemOpen
  \bibfield  {author} {\bibinfo {author} {\bibfnamefont {Z.}~\bibnamefont
  {Zhang}}, \bibinfo {author} {\bibfnamefont {A.}~\bibnamefont {Ahmadain}},\
  and\ \bibinfo {author} {\bibfnamefont {I.}~\bibnamefont {Klich}},\ }\bibfield
   {title} {\bibinfo {title} {Novel quantum phase transition from bounded to
  extensive entanglement},\ }\href
  {https://doi.org/https://doi.org/10.1073/pnas.1702029114} {\bibfield
  {journal} {\bibinfo  {journal} {Proceedings of the National Academy of
  Sciences}\ }\textbf {\bibinfo {volume} {114}},\ \bibinfo {pages} {5142}
  (\bibinfo {year} {2017})}\BibitemShut {NoStop}%
\bibitem [{\citenamefont {Crisanti}\ \emph {et~al.}(2012)\citenamefont
  {Crisanti}, \citenamefont {Paladin},\ and\ \citenamefont
  {Vulpiani}}]{crisanti2012products}%
  \BibitemOpen
  \bibfield  {author} {\bibinfo {author} {\bibfnamefont {A.}~\bibnamefont
  {Crisanti}}, \bibinfo {author} {\bibfnamefont {G.}~\bibnamefont {Paladin}},\
  and\ \bibinfo {author} {\bibfnamefont {A.}~\bibnamefont {Vulpiani}},\
  }\href@noop {} {\emph {\bibinfo {title} {Products of random matrices in
  Statistical Physics}}},\ Vol.\ \bibinfo {volume} {104}\ (\bibinfo
  {publisher} {Springer Science \& Business Media},\ \bibinfo {year}
  {2012})\BibitemShut {NoStop}%
\bibitem [{\citenamefont {Mochizuki}\ and\ \citenamefont
  {Hamazaki}(2024)}]{mochizuki2024absorption}%
  \BibitemOpen
  \bibfield  {author} {\bibinfo {author} {\bibfnamefont {K.}~\bibnamefont
  {Mochizuki}}\ and\ \bibinfo {author} {\bibfnamefont {R.}~\bibnamefont
  {Hamazaki}},\ }\bibfield  {title} {\bibinfo {title} {Absorption to
  fluctuating bunching states in nonunitary boson dynamics},\ }\href
  {https://doi.org/10.1103/PhysRevResearch.6.013004} {\bibfield  {journal}
  {\bibinfo  {journal} {Phys. Rev. Res.}\ }\textbf {\bibinfo {volume} {6}},\
  \bibinfo {pages} {013004} (\bibinfo {year} {2024})}\BibitemShut {NoStop}%
\end{thebibliography}%

\onecolumngrid
\newpage
\section*{Supplemental Material}
\twocolumngrid
\appendix

\section{How to realize generalized measurements}
\label{sec:generalized_measurement}
Generalized measurements on a system are realized in a way explained below. 
First, we prepare an ancilla, which is decoupled with the system. 
If the initial states of the system and ancilla are respectively $\ket{\psi_\mathrm{s}}$ and $\ket{\psi_\mathrm{a}}$, the initial state of the whole system is
\begin{align}
    \ket{\psi_\mathrm{w}}
    =\ket{\psi_\mathrm{s}}\otimes\ket{\psi_\mathrm{a}}.
\end{align}
Second, the system and ancilla are coupled by a unitary evolution $u$, 
\begin{align}
    \ket{\psi_\mathrm{w}}\rightarrow
    u\ket{\psi_\mathrm{w}}.
\end{align}
Third, we carry out the projective measurement on the ancilla. 
Here, we write the measurement basis as $\ket{\phi_\mathrm{a}^\omega}$, where $\{\omega\}$ are measurement outcomes. 
The probability that the measurement outcome becomes $\omega$ is
\begin{align}
    p(\omega)=\left|\bra{\phi_\mathrm{a}^\omega}
    u\ket{\psi_\mathrm{w}}\right|^2=\bra{\psi_\mathrm{s}}
    \mathsf{M}^\dagger_\omega \mathsf{M}_\omega\ket{\psi_\mathrm{s}},
\end{align}
where the Kraus operator $\mathsf{M}_\omega$ is
\begin{align}
    \mathsf{M}_\omega
    =\bra{\phi_\mathrm{a}^\omega}u\ket{\psi_\mathrm{a}}.
\end{align}
If we obtain one outcome $\omega$, the state of the system becomes
\begin{align}
    \ket{\psi_\mathrm{s}}\rightarrow
    \frac{\mathsf{M}_\omega\ket{\psi_\mathrm{s}}}
    {\sqrt{\bra{\psi_\mathrm{s}}\mathsf{M}^\dagger_\omega
    \mathsf{M}_\omega\ket{\psi_\mathrm{s}}}}.
\end{align}

To obtain one-qubit Kraus operators $\mathsf{M}_\pm(\eta)$ in the main text, we take a qubit as the ancilla, and choose the initial state as
\begin{align}
    \ket{\psi_\mathrm{a}}=\frac{1}{\sqrt{2}}
    (\ket{\uparrow_\mathrm{a}}+\ket{\downarrow_\mathrm{a}}). 
\end{align}
We also take the unitary evolution as
\begin{align}
    u(\eta)=\exp\left[i\theta(\eta)\sigma_3\otimes\sigma_2\right],
\end{align}
where 
\begin{align}
    \cos[\theta(\eta)]=\frac{1}{\sqrt{1+\eta^2}},\ \  
    \sin[\theta(\eta)]=\frac{\eta}{\sqrt{1+\eta^2}}.
\end{align}
Then, if we consider the projective measurement on the spin-$z$ component on the ancilla with $\ket{\phi_\mathrm{a}^{\omega=+}}=\ket{\uparrow_\mathrm{a}}$ and $\ket{\phi_\mathrm{a}^{\omega=-}}=\ket{\downarrow_\mathrm{a}}$, the Kraus operators become
\begin{align}
    \mathsf{M}_+(\eta)&
    =\bra{\uparrow_\mathrm{a}}u(\eta)\ket{\psi_\mathrm{a}}
    =\frac{\sigma_0+\eta\sigma_3}{\sqrt{2(1+\eta^2)}},\\
    \mathsf{M}_-(\eta)&
    =\bra{\downarrow_\mathrm{a}}u(\eta)\ket{\psi_\mathrm{a}}
    =\frac{\sigma_0-\eta\sigma_3}{\sqrt{2(1+\eta^2)}}.
\end{align}

\section{Monitored dynamics of pure states}
\label{sec:dynamics_pure-state}
We give details for the monitored dynamics of pure states exposed to generalized measurements.  
When local Haar-random unitaries are applied at time $t$, the pure state $\ket{\psi_t(\eta)}$ is transformed as
\begin{align}
    \ket{\psi_t(\eta)}\rightarrow
    U_t\ket{\psi_t(\eta)}.
\end{align}
After that, generalized measurements for all sites are carried out. 
When the measurement is done at a position $\ell$, the probability that the outcome becomes $\pm$ is determined through the Born rule,
\begin{align}
    p_{t,\ell}(\pm,\eta)=\bra{\psi_t(\eta)}M_{\pm,\ell}^\dagger(\eta)
    M_{\pm,\ell}(\eta)\ket{\psi_t(\eta)},
\end{align}
where $M_{\pm,\ell}(\eta)=\left(\bigotimes_{m=1}^{\ell-1}\sigma_0\right)\otimes\mathsf{M}_{\pm}(\eta)\otimes\left(\bigotimes_{m=\ell+1}^L\sigma_0\right)$. 
After the measurement at $\ell$ with the probabilistically determined outcome $\pm$, the state becomes
\begin{align}
    \ket{\psi_t(\eta)}\rightarrow
    \frac{M_{\pm,\ell}(\eta)\ket{\psi_t(\eta)}}
    {\sqrt{p_{t,\ell}(\pm,\eta)}}.
\end{align}
Thus, through one-step time evolution, $\ket{\psi_t(\eta)}$ is transformed as
\begin{align}
    \ket{\psi_t(\eta)}&\rightarrow\ket{\psi_{t+1}(\eta)}
    \nonumber\\&=\frac{M_t(\eta)U_t\ket{\psi_t(\eta)}}
    {\sqrt{\bra{\psi_t(\eta)}U^\dagger_t 
    M_t^\dagger(\eta)M_t(\eta)
    U_t\ket{\psi_t(\eta)}}},
\end{align}
where $M_t(\eta)=\prod_{\ell=1}^LM_{\omega_{t,\ell},\ell}(\eta)$. 
To obtain results in the main text, we randomly prepare an initial state $\ket{\psi_0}$ and simulate the dynamics explained above through the Monte-Carlo sampling. 
Applying a specific Kraus operator $M_{\omega_{t,\ell},\ell}(\eta)$ to $\ket{\psi_t(\eta)}$ corresponds to the postselection where $\omega_{t,\ell}$ is selected among various random measurement outcomes. 
In general, such postselections make experiments unscalable since the probability of sampling one specific trajectory becomes quite small when postselections are carried out many times. 
In our setting where measurements are done at all sites, which are also considered before \cite{li2019measurement}, experiments are more difficult than those in the settings where measurements are done at randomly selected sites~  \cite{li2018quantum,
skinner2019measurement,
li2019measurement}. 

\section{Numerical procedure for computing Lyapunov exponents}
\label{sec:numerics_Lyapunov-exponenets}
We can compute the Lyapunov spectrum of noisy dynamics by focusing on time evolutions of state vectors \cite{crisanti2012products,
ershov1998concept} as considered in Refs. \cite{zabalo2022operator,
kumar2024boundary,
mochizuki2024absorption,
aziz2024critical}. 
First, we choose $n$ initial states $\{\ket{\psi_0^i}\}$ randomly, with $i=1,2,\cdots,n$. 
We consider updating the state $\ket{\phi_{s}^i}=\ket{\psi_{t=sb}^i}$ by using the Grami-Schmidt orthonormalization procedure as follows:
\begin{align}
    \ket{\varphi_{s+1}^i}
    &=\tilde{V}_{s+1,b}\ket{\phi_s^i},\\
    \ket{\chi_{s+1}^i}
    &=(I_N-\Pi_{s+1}^i)
    \ket{\varphi_{s+1}^i},\\     
    \ket{\phi_{s+1}^i}
    &=\ket{\chi_{s+1}^i}
    /\sqrt{\langle\chi_{s+1}^i
    |\chi_{s+1}^i\rangle},
  \label{eq:dynamics}
\end{align}
where $I_N$ is the $N \times N$ identity operator with $N=2^L$ and $\tilde{V}_{s,b}$ is the time-evolution operator from $t=(s-1)b$ to $t=sb$ with $s$ and $b$ being positive integers. 
Here, $\Pi_s^i=\sum_{j=1}^{i-1}\ket{\phi_s^j}\bra{\phi_s^j}$ with $i\geq2$ is the projection operator onto the Hilbert space spanned by $\{\ket{\phi_s^j}\}_{j=1}^{i-1}$ and $\Pi_s^i=0$ for $i=1$.

Every time the dynamics proceed $b$ steps, we compute Lyapunov exponents at $t=sb$ by evaluating decay rates of the evolution from $\{\ket{\phi_s^i}\}$ to $\{\ket{\varphi_{s+1}^i}\}$. 
At $t=sb$, the sum of the Lyapunov exponents $\tilde{\varepsilon}_{sb,i}=\sum_{j=1}^i\varepsilon_{sb,j}$ is obtained through
\begin{align}
    \tilde{\varepsilon}_{sb,i}
    =-\frac{1}{sb}
    \sum_{r=1}^s\ln[\mathrm{VOL}
    (\varphi_r^1,\varphi_r^2,
    \cdots,\varphi_r^i)],
    \label{eq:Lyapunov-exponenents_sum}
\end{align}
where $\mathrm{VOL}(\varphi_r^1,\varphi_r^2,\cdots,\varphi_r^i)=\sqrt{\det\left(\Phi^\dagger_{r,i}\Phi_{r,i}\right)}$ is the volume of the space spanned by $\{\ket{\varphi_r^j}\}_{j=1}^i$, with $\Phi_{r,i}=(\ket{\varphi_r^1},\ket{\varphi_r^2},\cdots,\ket{\varphi_r^i})$. 
Then, $\tilde{\varepsilon}_i=\sum_{j=1}^i\varepsilon_j$ becomes
\begin{align}
    \tilde{\varepsilon}_i
    =\lim_{s\rightarrow\infty}
    \tilde{\varepsilon}_{sb,i}.
    \label{eq:lyapunov-spectrum_sum}
\end{align}
We can easily obtain $\{\varepsilon_i\}$ and $\{\varepsilon_{t,i}\}$ from $\{\tilde{\varepsilon}_i\}$ and $\{\tilde{\varepsilon}_{t,i}\}$. 
We can also obtain $\{\varepsilon_{sb,i}\}$ through
\begin{align}
    \varepsilon_{sb,i}=-\frac{1}{sb}\sum_{r=1}^s
    \ln\left(\sqrt{\langle\chi_r^i|\chi_r^i\rangle}\right).
    \label{eq:lyapunov-spectrum_each}
\end{align}
Both Lyapunov spectra obtained from Eqs. (\ref{eq:lyapunov-spectrum_sum}) and (\ref{eq:lyapunov-spectrum_each}) asymptotically take the same values as that computed through the singular values of $V_{t=sb}=\tilde{V}_{s,b}\tilde{V}_{s-1,b}\cdots\tilde{V}_{2,b}\tilde{V}_{1,b}$ \cite{ershov1998concept}. 
The effective Hamiltonian can be constructed from the whole Lyapunov spectrum and the Gram-Schmidt vectors,
\begin{align}
    K_t=\sum_{i=1}^N\varepsilon_{t,i}
    \ket{\phi_t^i}\bra{\phi_t^i}.
    \label{eq:effective-Hamiltonian}
\end{align}

\begin{figure}[tbp]
\begin{center}
\includegraphics[width=7.5cm]{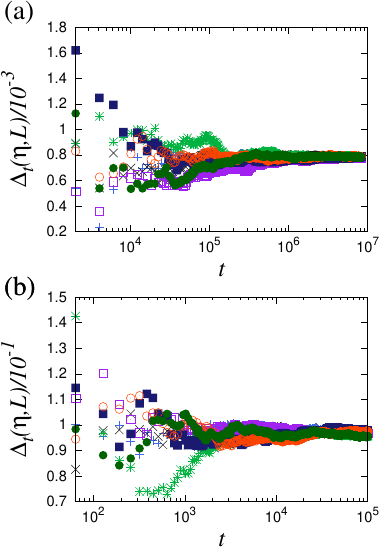}
\caption{Spectral gaps $\Delta_t(\eta,L)$ as functions of $t$ with (a) $\eta=0.1$ in the gapless phase and (b) $\eta=0.3$ in the gapped phase. 
The different symbols correspond to different trajectories starting from randomly chosen initial states. 
The parameter $b$ is taken to be $b=2048$ in (a) and $b=64$ in (b). 
The system size is $L=12$ in both (a) and (b).}
\label{fig:gap_t-dependence}
\end{center}
\end{figure}

We apply the abovementioned method to our quantum dynamics exposed to the generalized measurements. 
This is because computing Lyapunov exponents through the exact diagonalization of  $V_t(\eta)V^\dagger_t(\eta)$ is difficult to carry out in the long-time regime, due to (i) the exponential decay of singular values $\{\Lambda_{t,i}(\eta,L)\}$, which makes $\{\Lambda_{t,i}(\eta,L)\}$ deviate from the numerical precision, and (ii) the longer computational time to diagonalize $V_t(\eta)$ whose matrix size is exponentially large with respect to $L$. 
The Kraus operators $\{M_t(\eta)\}$ included in $\tilde{V}_{s,b}(\eta)=M_{sb}(\eta)U_{sb}M_{sb-1}(\eta) \cdots U_{(s-1)b+2}M_{(s-1)b+1}(\eta)U_{(s-1)b+1}$, which are independent of $i$, are generated through the Born rule based on $\ket{\psi_t^1(\eta)}$.  
Figure \ref{fig:gap_t-dependence} shows the spectral gaps $\Delta_t(\eta,L)$ computed through various trajectories. 
We can understand that the gaps converge to a trajectory-independent value in both gapless and gapped phases. 
Likewise, we assume that $\{\varepsilon_{t,i}(\eta,L)\}$ converge to trajectory-independent values. 
Then, we compute the averages of $\{\varepsilon_{t,i}(\eta,L)\}$ over $c$ steps from $t-bc$ to $t$ through one trajectory, 
\begin{align}
\left<\varepsilon_{t,i}(\eta,L)\right>
=\frac{1}{c}\sum_{\tilde{c}=0}^{c-1}
\varepsilon_{t-b\tilde{c},i}(\eta,L),
\end{align}
where the value of $c$ is taken in the range $100 \leq c \leq 1000$. 
When $\delta\varepsilon_{t,i}(\eta,L)/\left<\varepsilon_{t,i}(\eta,L)\right>$ becomes smaller than some threshold $d$, we stop the numerical simulation and adopt the average as the $i$th Lyapunov exponent $\varepsilon_i(\eta,L)=\left<\varepsilon_{t,i}(\eta,L)\right>$, where $\delta\varepsilon_{t,i}(\eta,L)$ is the standard deviation of $\varepsilon_{t,i}(\eta,L)$ with respect to time. 
Here, we take $d$ smaller than $3\times10^{-2}$.

We note that the numerical results depend on $b$ when the value of $b$ is too small.  
Thus, we explain how to choose the appropriate $b$ below. 
First, for the system size $L=6 \times m_L$ and the measurement strength $\eta=0.1 \times m_\eta$, we compute $\Delta(\eta,L)$, where $m_L=1,2,3$ and $m_\eta=1,2,3,4,5,6,7$. 
Second, we increase $b$ from some small value and compute the spectral gap for each $b$, with $m_L,\,m_\eta$, and $c$ being fixed. 
If the Lyapunov exponents become independent of $b$ for $b \geq b_\mathrm{min}(m_L,m_\eta,c)$ within the precision of $2d$, we use $b$ larger than $b_\mathrm{min}(m_L,m_\eta,c)$ in the range $6 \times m_L \leq L < 6\times(m_L+1)$ and $0.1 \times (m_\eta-1) < \eta \leq 0.1 \times m_\eta$. 
When we only compute some low-lying Lyapunov spectrum with small $i \ll N$, we can reach a sufficiently large $b$ following the procedure explained above. 
Meanwhile, when we compute the whole Lyapunov spectrum and the effective Hamiltonian, it becomes difficult to make $b$ large such that results become independent of $b$. 
This is because $\{\varepsilon_i(\eta,L)\}$ for large $i$ lead to fast exponential decay of corresponding Lyapunov vectors and thus the amplitudes of $\{\ket{\chi_{s,i}(\eta)}\}$ can easily deviate from the numerical precision for large $b$. 
Therefore, $b$ used in End Matter to obtain effective Hamiltonains through Eq. (\ref{eq:effective-Hamiltonian}) is not large enough for the results to be $b$-independent. 
However, we here assume that the effective Hamiltonians computed from small $b$ qualitatively capture features of true effective Hamiltonians, which should be independent of $b$. 
This is because we can confirm that behaviors of $W_{t,i}^r(\eta)$ become qualitatively same in cases of $b=512$ and $b=2048$ with $\eta=0.01$ and $L=10$. 

\begin{figure}
\begin{center}
\includegraphics[width=7cm]{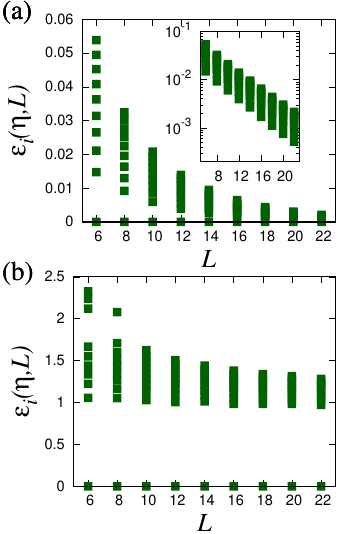}
\caption{Lyapunov spectra as functions of the system size $L$ up to $i=10$, with measurement strengths (a) $\eta=0.1$ in the gapless phase and (b) $\eta=0.7$ in the gapped phase. 
The inset in (a) is the semi-log plot of the data, showing that the low-lying Lyapunov spectrum exponentially decreases as $L$ is increased. }
\label{fig:spectrum_L-dependence}
\end{center}
\end{figure}

Figure \ref{fig:spectrum_L-dependence} shows the Lyapunov spectra up to $i=10$ as functions of the system size $L$, obtained along the abovementioned scheme. 
Note that $\{\varepsilon_i(\eta,L)\}$ are shifted such that $\varepsilon_1(\eta,L)=0$ is satisfied. 
The spectral gap $\Delta(\eta,L)$ with $\eta=0.1$ exponentially decreases as $L$ increases, while $\Delta(\eta,L)$ with $\eta=0.7$ is almost independent of $L$. 
The low-lying excited spectrum $\{\varepsilon_i(\eta=0.1,L)\}$ with $i\geq2$ also exhibits the exponential decay with respect to $L$, which is different from that in the gapless phase of isolated quantum systems. 

\section{Extrapolation of the spectral gap}
\label{sec:extrapolation_gap}
\begin{figure}[tbp]
\begin{center}
\includegraphics[width=7.5cm]{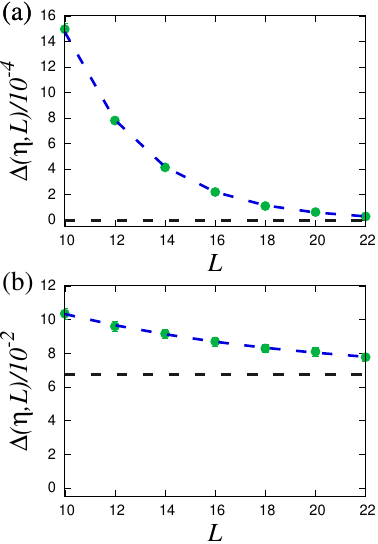}
\caption{The spectral gaps $\Delta(\eta,L)$ as functions of $L$, with (a) $\eta=0.1$ in the gapless phase and (b) $\eta=0.3$ in the gapped phase. 
The green circles are $\Delta(\eta,L)$, the blue broken lines represent the fitting curve $\Delta_\mathrm{fit}[\Gamma(\eta),L]$, and the black broken lines are $\Delta(\eta)$. 
The fitting parameters are (a) $\Delta(\eta)\simeq-5.2\times10^{-6},\,\alpha(\eta)\simeq1.4,\,\beta(\eta)\simeq3.4\times10^{-2}$ and (b) $\Delta(\eta)\simeq6.7\times10^{-2},\,\alpha(\eta)\simeq1.1,\,\beta(\eta)\simeq9.9\times10^{-2}$.}
\label{fig:gap_L-dependence}
\end{center}
\end{figure}

To obtain the spectral gap in the thermodynamic limit,
\begin{align}
    \Delta(\eta)
    =\lim_{L\rightarrow\infty}\Delta(\eta,L),
\end{align}
we extrapolate the numerical data $\Delta(\eta,L)$ in system sizes $10 \leq L \leq 22$. 
We use the least-squares method with the fitting function
\begin{align}
    \Delta_\mathrm{fit}[C(\eta),L]
    =D(\eta)+a(\eta)[b(\eta)]^{-L},
\end{align}
where $C(\eta)=[D(\eta),a(\eta),b(\eta)]$. 
Here, the function minimized is
\begin{align}
    \Theta[C(\eta)]=\sum_{L=10}^{22}
    \frac{\left(\Delta(\eta,L)-\Delta_\mathrm{fit}[C(\eta),L]\right)^2}{f^2(\eta,L)},
\end{align}
where $f(\eta,L)=d\times\Delta(\eta,L)$. 
For each $\eta$, we adopt $\Delta(\eta)$ as the spectral gap in the thermodynamic limit, such that 
\begin{align}
\Theta[\Gamma(\eta)]=\min_{C(\eta)}\Theta[C(\eta)]
\end{align} 
is satisfied with $\Gamma(\eta)=[\Delta(\eta),\alpha(\eta),\beta(\eta)]$. 
For the minimization of $\Theta[C(\eta)]$, we explore the range of $D(\eta)$, from $-\min_L[\Delta(\eta,L)]$ to $+\min_L[\Delta(\eta,L)]$. 
Figure \ref{fig:gap_L-dependence} shows the system-size dependence of the spectral gap, where the fitting curves $\Delta_\mathrm{fit}[\Gamma(\eta),L]$ (blue broken line) and numerical data $\Delta(\eta,L)$ (green circles) agree well. 
As shown in Fig. \ref{fig:gap_L-dependence}, $\Delta(\eta,L)$ converges to $\Delta(\eta)\simeq0$ for small $\eta$ corresponding to the gapless phase, while $\Delta(\eta)$ is far from $0$ for large $\eta$ in the gapped phase. 

\begin{figure}[tbp]
\begin{center}
\includegraphics[width=7.5cm]{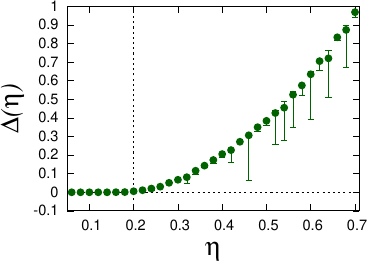}
\caption{The spectral gap $\Delta(\eta)$ as a function of $\eta$, which is obtained through the least-squares method.}
\label{fig:gap_e-dependence}
\end{center}
\end{figure}

After obtaining $\Delta(\eta)$, we compute the error bar of $\Delta(\eta)$ following the procedure explained below. 
We again sweep $D(\eta)$, $a(\eta)$, and $b(\eta)$ in the range $-\min_L[\Delta(\eta,L)]<D(\eta)<+\min_L[\Delta(\eta,L)]$, $0<a(\eta)<2\alpha(\eta)$, and $0<b(\eta)<2\beta(\eta)$, respectively. 
Then, we evaluate
\begin{align}
    \tilde{\Theta}[C(\eta)]
    =\sum_{L=10}^{22}
    \frac{\left(\Delta_\mathrm{fit}[C(\eta),L]
    -\Delta_\mathrm{fit}[\Gamma(\eta),L]\right)^2}{f^2(\eta,L)}.
\end{align}
We obtain error bars of $\Delta(\eta)$ through $\min_{C(\eta)}D(\eta)$ and $\max_{C(\eta)}D(\eta)$ under the condition that $\tilde{\Theta}[C(\eta)]\leq\Theta[\Gamma(\eta)]$ is satisfied. 
Figure \ref{fig:gap_e-dependence} shows $\Delta(\eta)$ and error bars obtained through the abovementioned method. 
We consider that the system with the measurement strength $\eta$ is in the gapless phase if $\min_{C(\eta)}D(\eta)<0$ is satisfied, while $\min_{C(\eta)}D(\eta)>0$ corresponds to the gapped phase. 
In the range $\eta<\eta_c=0.2$, our numerical data satisfy the condition for the gapless phase.

\section{Spectral gap of measurement-only dynamics}
\label{sec:gap_measurement-only}
Under some assumptions, we can analytically show that the spectral gap takes a non-zero value independent of the system size when unitary gates are absent. 
The final result is in Eq. (\ref{eq:gap_no-unitary}). 
In such a situation, all qubits are independent, and thus we can write $V_t(\eta)$ as
\begin{align}
    V_t(\eta)&=\bigotimes_{\ell=1}^L
    \mathsf{V}_{t,\ell}(\eta),
    \label{eq:Vt_product}\\
    \mathsf{V}_{t,\ell}(\eta)
    &=\prod_{u=1}^t
    \mathsf{M}_{\omega_{u,\ell}}(\eta),
\end{align}
where $\mathsf{V}_{t,\ell}(\eta)$ is the $2\times2$ time-evolution operator for the site $\ell$, and $\omega_{u,\ell}=\pm1$ is probabilistically determined through the Born rule. 
The spectral gap and the Born probability are determined through the operators
\begin{align}
    \mathsf{M}_+^\dagger(\eta)
    \mathsf{M}_+(\eta)&=
    \left(\begin{array}{cc}
      p_\eta & 0 \\
      0 & 1-p_\eta
    \end{array}\right),\\
    \mathsf{M}_-^\dagger(\eta)
    \mathsf{M}_-(\eta)&=
    \left(\begin{array}{cc}
      1-p_\eta & 0 \\
      0 & p_\eta
    \end{array}\right),
\end{align}
where
\begin{align}
    p_\eta=\frac{(1+\eta)^2}{2(1+\eta^2)}\geq \frac{1}{2}.
\end{align}
In the following, we consider the case where $\eta\neq0$ and thus $p_\eta\neq1/2$ since $\eta=0$ corresponds to no measurement.  
If $\omega_{u,\ell}=+1$ is realized $v$ times from $u=1$ to $u=t$ and $2v \geq t$, the singular values of $\mathsf{V}_{t,\ell}(\eta)$ become
\begin{align}
    \Lambda_{t,\ell,v,1}(\eta)
    &=\sqrt{p_\eta^v(1-p_\eta)^{t-v}},\\
    \Lambda_{t,\ell,v,2}(\eta)
    &=\sqrt{p_\eta^{t-v}(1-p_\eta)^v}.
\end{align}
In this case, the spectral gap $\tilde{\Delta}_{t,\ell,v}(\eta)$ can be written as
\begin{align}
    \tilde{\Delta}_{t,\ell,v}(\eta)
    &=\left|\frac{1}{t}
    \ln\left[\frac{\Lambda_{t,\ell,v,1}(\eta)}{\Lambda_{t,\ell,v,2}(\eta)}\right]\right|
    \nonumber\\
    &=\frac{|t-2v|}{2t}\ln\left(\frac{p_\eta}{1-p_\eta}\right). 
    \label{eq:gap_no-unitary_v}
\end{align}
We can easily check that the gap becomes the same even when $2v \leq t$. 

Assuming a product initial state, $\ket{\psi_0}=\bigotimes_{\ell=1}^L\ket{\psi_{0,\ell}}$ with $\ket{\psi_{0,\ell}}=\binom{\psi_{0,\ell,\uparrow}}{\psi_{0,\ell,\downarrow}}$, we find that the Born probability that $\omega_{u,\ell}=+1$ is realized $v$ times becomes 
\begin{align}
    &\bra{\psi_{0,\ell}}
    \mathsf{V}_{t,\ell}^\dagger(\eta)
    \mathsf{V}_{t,\ell}(\eta)
    \ket{\psi_{0,\ell}}\nonumber\\
    &=\binom{t}{v}\left[p_\eta^v
    (1-p_\eta)^{t-v} |\psi_{0,\ell,\uparrow}|^2
    +p_\eta^{t-v}(1-p_\eta)^v
    |\psi_{0,\ell,\downarrow}|^2\right].
\end{align}
Therefore, the ensemble average of the spectral gap at time step $t$ can be written as
\begin{align}
    \left<\tilde{\Delta}_{t,\ell,v}(\eta)\right>
    =\sum_{v=0}^t
    \binom{t}{v}p_\eta^v
    (1-p_\eta)^{t-v}
    \tilde{\Delta}_{t,\ell,v}(\eta)
    \label{eq:gap_average_no-unitary}
\end{align}
Thus, the probability that the spectral gap is $\tilde{\Delta}_{t,\ell,v}(\eta)$ becomes the bimodal distribution. 
The mean value of $v$ becomes $v^*_{t,\eta}=tp_\eta$, which leads to
\begin{align}
    \tilde{\Delta}_{t,\ell,v^*_{t,\eta}}(\eta)
    =\tilde{\Delta}^*(\eta)
    :=\left(p_\eta-\frac{1}{2}\right)
    \ln\left(\frac{p_\eta}{1-p_\eta}\right).
\end{align}
When $t$ is sufficiently large, we can show that almost all trajectories exhibit $\tilde{\Delta}_{t,\ell,v}(\eta)=\tilde{\Delta}^*(\eta)$. 
To this end, using $p_\eta>1/2$ and thus $v^*_{t,\eta}>t/2$, we evaluate the probability that $\left|\tilde{\Delta}_{t,\ell,v}(\eta)-\tilde{\Delta}^*(\eta)\right|   >\theta\ln\left(\frac{p_\eta}{1-p_\eta}\right)$ is satisfied,
\begin{align}
    &P\left[\left|
    \tilde{\Delta}_{t,\ell,v}(\eta)
    -\tilde{\Delta}^*(\eta)\right|
    >\theta\ln\left(
    \frac{p_\eta}{1-p_\eta}
    \right)\right]\nonumber\\
    &=P\left[\left|\left|v-\frac{t}{2}\right|
    -\left|v^*_{t,\eta}-\frac{t}{2}
    \right|\right|>t\theta\right]\nonumber\\
    &=P\left[\left|v-v_{t,\eta}^*\right|
    >t\theta,\,v>\frac{t}{2}\right]\nonumber\\
    &\ \ +P\left[\left|t-(v+v_{t,\eta}^*)
    \right|>t\theta,\,v<\frac{t}{2}\right],
    \label{eq:probability_gap}
\end{align}
where $\theta$ is a positive value and $P\left[\left|v-v_{t,\eta}^*\right|>t\theta,\,v>\frac{t}{2}\right]$ is the joint probability that the conditions $\left|v-v_{t,\eta}^*\right|>t\theta$ and $v>\frac{t}{2}$ are satisfied. 
The probabilities in the right-hand side of Eq. (\ref{eq:probability_gap}) satisfy
\begin{align}
    &P\left[\left|v-v_{t,\eta}^*\right|
    >t\theta,\,v>\frac{t}{2}\right]
    \nonumber\\
    &\leq P\left[\left|v_{t,\eta}^*-v
    \right|>t\theta\right],
    \label{eq:inequality_probability_1}\\
    &P\left[\left|t-(v+v_{t,\eta}^*)
    \right|>t\theta,\,v<\frac{t}{2}\right]
    \nonumber\\
    &\leq P\left[v<\frac{t}{2}\right]
    \leq P\left[\left|v^*_{t,\eta}-v\right|>v^*_{t,\eta}-\frac{t}{2}\right].
    \label{eq:inequality_probability_2}
\end{align}
Since the variance of the bimodal distribution becomes $\sigma_{t,\eta}^2=tp_\eta(1-p_\eta)$, Eqs. (\ref{eq:probability_gap})-(\ref{eq:inequality_probability_2}) and the 
Chebyshev inequality lead to
\begin{align}
    &P\left[\left|
    \tilde{\Delta}_{t,\ell,v}(\eta)
    -\tilde{\Delta}^*(\eta)\right|>\theta
    \ln\left(\frac{p_\eta}{1-p_\eta}
    \right)\right]\nonumber\\
    &\leq\frac{p_\eta(1-p_\eta)}{t\theta^2}
    +\frac{p_\eta(1-p_\eta)}{t\left(p_\eta-\frac{1}{2}\right)^2}.
    \label{eq:Chebyshev-inequality_gap}
\end{align}
Through Eq. (\ref{eq:Chebyshev-inequality_gap}), we can evaluate the probability that $\left|\tilde{\Delta}_{t,\ell,v}(\eta)-\tilde{\Delta}^*(\eta)\right|$ becomes small at all sites $\ell$:
\begin{align}
    &P\left[\left|\tilde{\Delta}_{t,\ell,v}(\eta)-\tilde{\Delta}^*(\eta)\right|\leq\theta
    \ln\left(\frac{p_\eta}{1-p_\eta}\right)
    \ \forall \ell\right]\nonumber\\
    &=\prod_{\ell=1}^L\left(1-P\left[\left|\tilde{\Delta}_{t,\ell,v}(\eta)
    -\tilde{\Delta}^*(\eta)\right|>\theta
    \ln\left(\frac{p_\eta}{1-p_\eta}
    \right)\right]\right)\nonumber\\
    &\geq\left[1-\frac{p_\eta(1-p_\eta)}{t\theta^2}-\frac{p_\eta(1-p_\eta)}{t\left(p_\eta-\frac{1}{2}\right)^2}\right]^L\nonumber\\
    &\geq1-L\left[\frac{p_\eta(1-p_\eta)}{t\theta^2}+\frac{p_\eta(1-p_\eta)}{t\left(p_\eta-\frac{1}{2}\right)^2}\right].
\end{align}
Therefore, if $t \gg Lp_\eta(1-p_\eta)/\theta^2$ and $t \gg Lp_\eta(1-p_\eta)/\left(p_\eta-\frac{1}{2}\right)^2$ are satisfied, the spectral gap typically becomes $\tilde{\Delta}^*(\eta)$ for almost all trajectories. 
If we choose $\theta=O(1/L)$ and $t=\Omega(L^4)$ is satisfied, the probability that the gap satisfies $\tilde{\Delta}^*(\eta)-O(1/L)<\tilde{\Delta}_{t,\ell,v}(\eta)<\tilde{\Delta}^*(\eta)+O(1/L)$ for all $\ell$ becomes $1-O(1/L)$. 
Then, for large $t$ where $\tilde{\Delta}_{t,\ell,v}(\eta)$ has converged to $\tilde{\Delta}^*(\eta)$, the time-evolution operator becomes
\begin{align}
    \mathsf{V}_{t,\ell}(\eta)\propto
    \left(\begin{array}{cc}
    1&0\\0&e^{-\Omega_{t,\ell}
    \tilde{\Delta}^*(\eta)t}
    \end{array}\right),
\end{align}
with $\Omega_{t,\ell}=\mathrm{sign}\left(\sum_{u=1}^t\omega_{u,\ell}\right)$. 
In such a situation, through Eq. (\ref{eq:Vt_product}), we can understand that the spectral gap of the whole system also becomes $\Delta(\eta)=\tilde{\Delta}^*(\eta)$:
\begin{align}
    {\Delta}(\eta)
    =\left(p_\eta-\frac{1}{2}\right)\ln
    \left(\frac{p_\eta}{1-p_\eta}\right).
    \label{eq:gap_no-unitary}
\end{align}
Equation (\ref{eq:gap_no-unitary}) means that the system always resides in the gapped phase. 
In other words, the measurement-induced spectral transition revealed in the main text is caused by the competition between unitary dynamics and quantum measurements.

\section{Evaluation of the spectral width based on the majorization}
\label{sec:spectral-width}
To evaluate the maximal width of the Lyapunov spectrum, we consider the majorization for arrays of real numbers $u=(u_1,u_2,\cdots,u_N)$, where $u_i \geq u_{i+1}$ is satisfied. 
The majorization of two arrays $u$ and $v$, \begin{align}
    u \prec v,
\end{align}
is defined as 
\begin{align}
    \sum_{i=1}^nu_i \leq \sum_{i=1}^nv_i,
\end{align}
where the equality is achieved for $n=N$. 
If $u$ and $v$ satisfy $u \prec v$, there exists a doubly stochastic matrix $W$, which satisfies 
\begin{align}
    u=Wv.
\end{align}
Here, a doubly stochastic matrix is defined as a non-negative matrix satisfying $\sum_iW_{ij}=\sum_iW_{ji}=1$ for arbitrary $j$. 

The log-majorization of two arrays $u$ and $v$, 
\begin{align}
    u \prec_\mathrm{log} v,
\end{align} 
is defined as
\begin{align}
    \prod_{i=1}^nu_i \leq \prod_{i=1}^nv_i, \end{align}
where the equality is satisfied for $n=N$. 
We consider the log-majorization for the singular values of matrices. 
Given a matrix $A$, we consider the array $s(A)=[s_1(A),s_2(A),\cdots,s_N(A)]$,  
where $\{s_i(A)\}$ denote the singular values of $A$ and $s_i(A) \geq s_{i+1}(A)$ is satisfied. 
The Horn theorem tells us that a product of two matrices satisfies
\begin{align}
    s(AB) \prec_\mathrm{log} s(A)s(B),
    \label{eq:Horn-theorem}
\end{align}
where $s(A)s(B)=[s_1(A)s_1(B),\cdots,s_N(A)s_N(B)]$ \cite{hiai2024log}. 

We apply the properties of the majorizations explained above to our Kraus operators. 
From Eq. (\ref{eq:Horn-theorem}), we can show that 
each Kraus operator satisfies
\begin{align}
    s[M_t(\eta)]\prec_\mathrm{log} 
    &s[\tilde{M}_{\omega_{t,\ell=1}}(\eta)]s[\tilde{M}_{\omega_{t,\ell=2}}(\eta)] \cdots s[\tilde{M}_{\omega_{t,\ell=L}}(\eta)]\nonumber\\
    &=\mathbb{M}(\eta),
    \label{eq:log-majorization_Mt}
\end{align}
where $\tilde{M}_{\omega_{t,\ell}}(\eta)=\left[\bigotimes_{m=1}^{\ell-1}\sigma_0\right]\otimes \mathsf{M}_{\omega_{t,\ell}}(\eta)\otimes\left[\bigotimes_{m=\ell+1}^L\sigma_0\right]$ with $\sigma_0$ being the $2\times2$ identity matrix. 
Here, the array $\mathbb{M}(\eta)$ is independent of $\{\omega_{t,\ell}\}$,
\begin{align}
    [\mathbb{M}(\eta)]_i
    =\begin{cases}
    \left[\frac{1+\eta}{\sqrt{2(1+\eta^2)}}\right]^L
    &(1 \leq i \leq N/2),\\
    \left[\frac{1-\eta}{\sqrt{2(1+\eta^2)}}\right]^L
    &(N/2 < i \leq N),
    \end{cases}
\end{align}
where $N=2^L$. 
This is because singular values of $\mathsf{M}_+(\eta)$ and $\mathsf{M}_-(\eta)$ are the same, $[(1+\eta)/\sqrt{2(1+\eta^2)},(1-\eta)/\sqrt{2(1+\eta^2)}]$. 
In a similar manner, from Eqs. (\ref{eq:Horn-theorem}) and (\ref{eq:log-majorization_Mt}), we can understand that our time-evolution operator $V_t(\eta)=M_t(\eta)U_t \cdots M_1(\eta)U_1$ satisfies 
\begin{align}
    s\left[V_t(\eta)\right] \prec_\mathrm{log} 
    s[M_1(\eta)]\cdots s[M_t(\eta)]
    \prec_\mathrm{log}\mathbb{M}_t(\eta),
    \label{eq:log-majorization_Vt}
\end{align}
where the array $\mathbb{M}_t(\eta)$, which is independent of measurement outcomes, is 
\begin{align}
    [\mathbb{M}_t(\eta)]_i
    =\begin{cases}
    \left[\frac{1+\eta}{\sqrt{2(1+\eta^2)}}\right]^{Lt}
    &(1 \leq i \leq N/2),\\
    \left[\frac{1-\eta}{\sqrt{2(1+\eta^2)}}\right]^{Lt}
    &(N/2 < i \leq N).
    \end{cases}
\end{align}
Note that the right-hand side of Eq. (\ref{eq:log-majorization_Vt}) is independent of unitary matrices $\{U_t\}$ because of $s_i(U_t)=1$. 
Equation (\ref{eq:log-majorization_Vt}) means that an array 
\begin{align}
    \mu_i(\eta,L)=\begin{cases}
    L\ln\left[\frac{1+\eta}{\sqrt{2(1+\eta^2)}}\right]
    &(1 \leq i \leq N/2),\\
    L\ln\left[\frac{1-\eta}{\sqrt{2(1+\eta^2)}}\right]
    &(N/2 < i \leq N)
    \end{cases}
\end{align}
majorizes $-\varepsilon(\eta,L)$,
\begin{align}
    -\varepsilon(\eta,L) \prec \mu(\eta,L),
\end{align}
where $-\varepsilon(\eta,L)=[-\varepsilon_1(\eta,L),\cdots,-\varepsilon_N(\eta,L)]$. 
This means that $-\varepsilon(\eta,L)$ and $\mu(\eta,L)$ are related by a doubly stochastic matrix $W(\eta,L)$, that is,
\begin{align}
    \varepsilon_i(\eta,L)
    =L&\left(w_i(\eta,L)
    \ln\left[\frac{\sqrt{2(1+\eta^2)}}{1+\eta}\right]
    \right.\nonumber\\
    &\left.+[1-w_i(\eta,L)]
    \ln\left[\frac{\sqrt{2(1+\eta^2)}}{1-\eta}\right]\right)
\end{align}
is satisfied with $w_i(\eta,L)=\sum_{j=1}^{N/2}W_{ij}(\eta,L)$ being real and non-negative numbers less than $1$. 
This results in 
\begin{align}
    &\varepsilon_N(\eta,L)-\varepsilon_1(\eta,L)\nonumber\\
    &=L[w_1(\eta,L)-w_N(\eta,L)]\ln\left(\frac{1+\eta}{1-\eta}\right).
\end{align}
Since the coefficients reside in the range $0 \leq w_i(\eta,L) \leq 1$, the upper bound of the spectral width becomes $O(L)$,
\begin{align}
    \varepsilon_N(\eta,L)-\varepsilon_1(\eta,L)
    \leq L\ln\left(\frac{1+\eta}{1-\eta}\right).
\end{align}
Thus, the typical level spacing becomes $\exp[-O(L)]$ since the number of $\{\varepsilon_i(\eta)\}$ is $N=2^L$.

\end{document}